\documentclass{aa}
\usepackage[varg]{txfonts}
\usepackage{epstopdf}
\usepackage{lscape}
\usepackage{color}

\usepackage{natbib,twoopt}

\usepackage[breaklinks=true]{hyperref} 
\bibpunct{(}{)}{;}{a}{}{,}             
\makeatletter
  \newcommandtwoopt{\citeads}[3][][]{\href{http://adsabs.harvard.edu/abs/#3}%
    {\def\hyper@linkstart##1##2{}%
     \let\hyper@linkend\@empty\citealp[#1][#2]{#3}}}
  \newcommandtwoopt{\citepads}[3][][]{\href{http://adsabs.harvard.edu/abs/#3}%
    {\def\hyper@linkstart##1##2{}%
     \let\hyper@linkend\@empty\citep[#1][#2]{#3}}}
  \newcommandtwoopt{\citetads}[3][][]{\href{http://adsabs.harvard.edu/abs/#3}%
    {\def\hyper@linkstart##1##2{}%
     \let\hyper@linkend\@empty\citet[#1][#2]{#3}}}
  \newcommandtwoopt{\citeyearads}[3][][]%
    {\href{http://adsabs.harvard.edu/abs/#3}
{\def\hyper@linkstart##1##2{}%
\let\hyper@linkend\@empty\citeyear[#1][#2]{#3}}}
\makeatother
\newcommand\kms{$\mbox{km s}^{-1}$}

\hypersetup{draft} 
\begin{document}

\title{The VLT-FLAMES Tarantula Survey}

\subtitle{XXX. Red stragglers in the clusters Hodge\,301 and SL\,639\thanks{Based on observations at the European Southern Observatory in programme 182.D-0222}
}

\author{N. Britavskiy\inst{1, 2}
  \and D. J. Lennon\inst{3, 1}
  \and L. R. Patrick\inst{1, 2}
  \and C. J. Evans\inst{4}
  \and A. Herrero\inst{1, 2}
  \and N. Langer\inst{5}
  \and  J. Th. van Loon\inst{6} \and \\
   J. S. Clark\inst{7}
  \and F. R. N. Schneider\inst{8,9,10}
  \and L. A. Almeida\inst{11}
  \and H. Sana\inst{12}
  \and A. de Koter\inst{13}
  \and W. D. Taylor\inst{4} 
}

\offprints{britavskiy@iac.es}
\authorrunning{N. Britavskiy et al.}  
\titlerunning{Physical parameters of RSGs in 30~Dor}

\institute{Instituto de Astrof\'isica de Canarias, E-38205 La Laguna, Tenerife, Spain 
  \and Universidad de La Laguna, Dpto. Astrof\'isica, E-38206 La Laguna, Tenerife, Spain
  \and ESA, European Space Astronomy Centre, Apdo. de Correos 78, E-28691 Villanueva de la Cañada, Madrid, Spain
\and UK Astronomy Technology Centre, Royal Observatory, Blackford Hill, Edinburgh, EH9 3HJ, UK
\and Argelander-Institut für Astronomie der Universität Bonn, Auf dem Hügel 71, 53121 Bonn, Germany
\and  Lennard-Jones Laboratories, Keele University, Staffordshire, ST5 5BG, United Kingdom
\and School of Physical Sciences, The Open University, Walton Hall, Milton Keynes, MK7 6AA, United Kingdom
\and Department of Physics, University of Oxford, Denys Wilkinson Building, Keble Road, Oxford OX1 3RH, United Kingdom
\and Zentrum f\"{u}r Astronomie der Universit\"{a}t Heidelberg, Astronomisches Rechen-Institut, M\"{o}nchhofstr. 12-14, 69120 Heidelberg, Germany
\and Heidelberger Institut f\"{u}r Theoretische Studien, Schloss-Wolfsbrunnenweg 35, 69118 Heidelberg, Germany
\and Departamento de F\'isica Te\'orica e Experimental, Universidade Federal do Rio Grande do Norte, CP 1641, Natal, RN, 59072-970, Brazil
\and Institute of astrophysics, KU Leuven, Celestijnlaan 200D, 3001, Leuven, Belgium
\and Anton Pannenkoek Institute for Astronomy, University of Amsterdam, 1090 GE Amsterdam, The Netherlands}

\date{Received 2 November 2018 / Accepted 26 February 2019}

\abstract{}
{We estimate physical parameters for the late-type massive stars observed as part of the VLT-FLAMES Tarantula Survey (VFTS) in the 30~Doradus region of the Large Magellanic Cloud (LMC).}
{The observational sample comprises 20 candidate red supergiants (RSGs) which are the reddest (($B-V$)\,$>$\,1 mag) and brightest ($V$\,$<$\,16 mag) objects in the VFTS. We use optical and near-IR photometry to estimate their temperatures and luminosities, and introduce the luminosity--age diagram to estimate their ages.}
{We derive physical parameters for our targets, including temperatures from a new calibration of $(J-K_{\rm s})_{0}$ colour for luminous cool stars in the LMC, luminosities from their $J$-band magnitudes (thence radii), and ages from comparisons with state-of-the-art evolutionary models.
We show that interstellar extinction is a significant factor for our targets, highlighting the need to take it into account in analysis of the physical parameters of RSGs. We find that some of the candidate RSGs could be massive AGB stars. The apparent ages of the RSGs in the Hodge~301 and SL\,639 clusters show a significant spread (12--24\,Myr). We also apply our approach to the RSG population of the relatively nearby NGC\,2100 cluster, finding a similarly large spread.}
{We argue that the effects of mass-transfer in binaries may lead to more massive and luminous RSGs (which we call `red stragglers') than expected from single-star evolution, and that the true cluster ages correspond to the upper limit of the estimated RSG ages. In this way, the RSGs can serve as a new and potentially reliable age tracer in young star clusters. The corresponding analysis yields ages of 24$^{+5}_{-3}$\,Myr for Hodge~301, 22$^{+6}_{-5}$\,Myr for SL\,639, and 23$^{+4}_{-2}$\,Myr for NGC\,2100.}
\maketitle

\section{Introduction}     \label{sec:introduction}

Multi-epoch spectroscopy of an unprecedented sample of hot, massive stars in the 30~Doradus region of the Large Magellanic Cloud (LMC) was obtained by the VLT-FLAMES Tarantula Survey \citep[VFTS,][hereafter Paper~I]{vfts}. 
To try to obtain an unbiased view of the massive-star population of 30~Dor, no restrictions on colour were employed in the VFTS target list, so as to 
potentially include heavily reddened O-type stars that were expected to be present in the region. The resulting observed sample therefore included spectra of 91 later-type stars in the region (with spectral types ranging from early A to M, see Table~3 of Paper~I), plus spectra for an additional 102 stars thought to be mostly cool foreground stars. Among these two sets of cool stars are $\sim$20 stars that are known red supergiants (RSGs) or new candidate RSGs. Here we investigate their stellar parameters and evolutionary status, and discuss their ages in the context of the age of the 30~Dor region and its component stellar groups.

RSGs represent the final evolutionary stage of most massive stars before core-collapse. They also represent the physically largest evolutionary phase possible for single stars, making them critical for understanding the total fractions, mass ranges, and evolutionary states of interacting massive binaries \citep{emily_book}. Indeed, the high incidence of interacting massive binaries \citep{Sana_2012} is expected to lead to about half of the type II supernova population resulting from post-interaction or merged stars \citep[][Zapartas et al. in prep.]{Podsiadlowski_1992} and to delayed supernovae relative to single star evolution \citep[e.g.][]{Zapartas_2017}. However, our understanding of these stars is hindered as correct determination of their physical parameters is still challenging due to the many uncertainties associated with modelling their complex atmospheres and winds \citep[e.g.][]{Massey2005, Levesque10_phys,Davies_2013}. The evolutionary history of RSGs also depends on metallicity, initial mass, and probably binarity. The RSGs found in two clusters in the 30~Doradus region, Hodge~301 \citep{Hodge_1988} and SL\,639 \citep{SL639}, are particularly interesting because they enable study of two samples of RSGs that are each presumably coeval, as discussed later.



The wavelength coverage of the VFTS spectra was tailored to the analysis of OB-type stars (see Paper~I). While useful to classify late-type stars, the coverage is not sufficient to estimate effective temperatures for RSGs, so we have resorted to photometric methods to investigate the physical properties of our sample. There are a number of different approaches discussed in the literature, so we briefly review these to assess potential advantages or drawbacks. A radial-velocity (RV) analysis of the VFTS spectroscopy of the sample is presented in a companion paper (Patrick et al. submitted). 

This paper is organised as follows: Section~\ref{sec:observations} introduces our observational sample and Section~\ref{analysis} uses three photometric techniques to estimate the physical parameters of each star. We discuss our results in Section~\ref{sec:discussion}, with brief conclusions presented in Section~\ref{sec:conclusion}.



\section{Observational sample}    \label{sec:observations}

\begin{figure}
\begin{center}
\includegraphics[width=9.5cm]{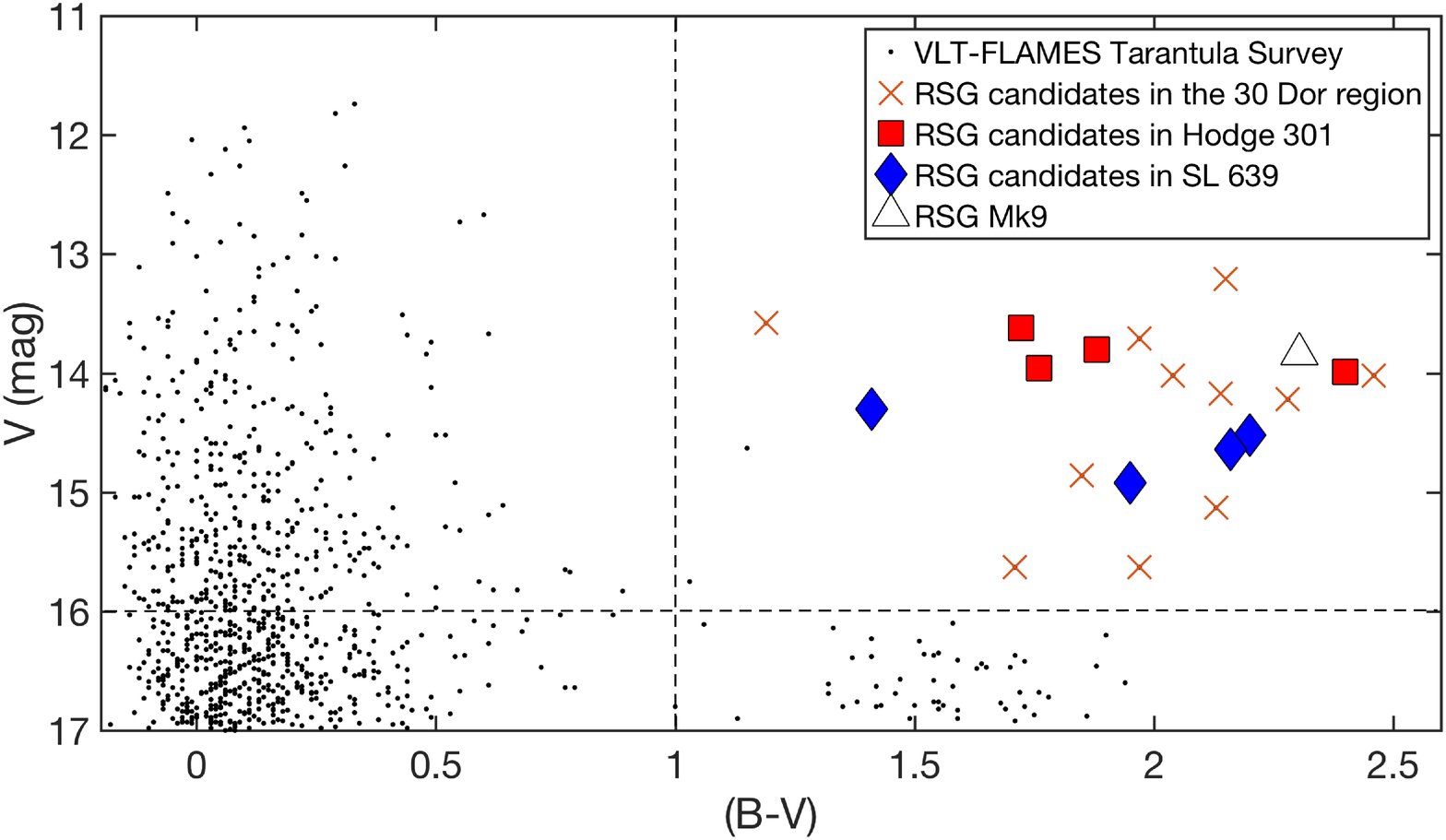}
\caption[]{Colour--magnitude diagram for the VFTS targets, highlighting the candidate RSGs studied and their membership of (or association with) the Hodge~301 and SL\,639 clusters. The photometric criteria ($V$\,$<$\,16\,mag, $(B-V)$\,$>$\,1\,mag) are indicated by the dashed lines.\label{cmd}}
\end{center}
\end{figure}

The VFTS included observations of 91 targets with spectral types ranging from A through to M (see Table 3 in Paper I). To select candidate RSGs for this study (and that of Patrick et al.) we used photometric criteria of ($B-V$)\,$>$\,1 mag and $V$\,$<$\,16 mag, as shown in the colour--magnitude diagram in Fig.~\ref{cmd}\footnote{The two bluest targets that satisfy these criteria were not considered further as they were classified by \citet{Ostars_2012} as heavily reddened O-type stars.}. 

Apparently foreground stars with RV\,$<$\,100\,km~s$^{-1}$ were excluded from the VFTS sample by inspection of the spectra at the outset of the project (see Section~2.2.2. of Paper~I). At that stage, we also omitted a small number of cool stars that appeared to have RVs consistent with membership of the LMC but with very low signal-to-noise (S/N) spectra.  In the context of this study and the RV analysis by Patrick et al. (submitted), these low S/N spectra can still provide new insights, so we supplemented the primary VFTS targets with three stars that were omitted from Paper~I but that appear to be members of the LMC. For future reference the full listing of the 102 previously discarded targets is given in Table~\ref{rejects}, where they are given IDs of the form {\it 2xxx} to distinguish them from the primary VFTS catalogue.


Given our focus on Hodge 301 and SL 639 we also include photometry of RSG WB97\#5 \citep{WB_1997,Grebel_2000} in Hodge~301, which was not observed in the VFTS due to crowding in the core of this cluster. For completeness, we also consider the brightest RSG in the region, Mk\,9 \citep{mk85}; by chance this was not included in the VFTS sample due to the high density of targets in and around R136 (which limited the fibre allocations). 

\begin{figure*}
\begin{center}
\includegraphics[scale=0.6]{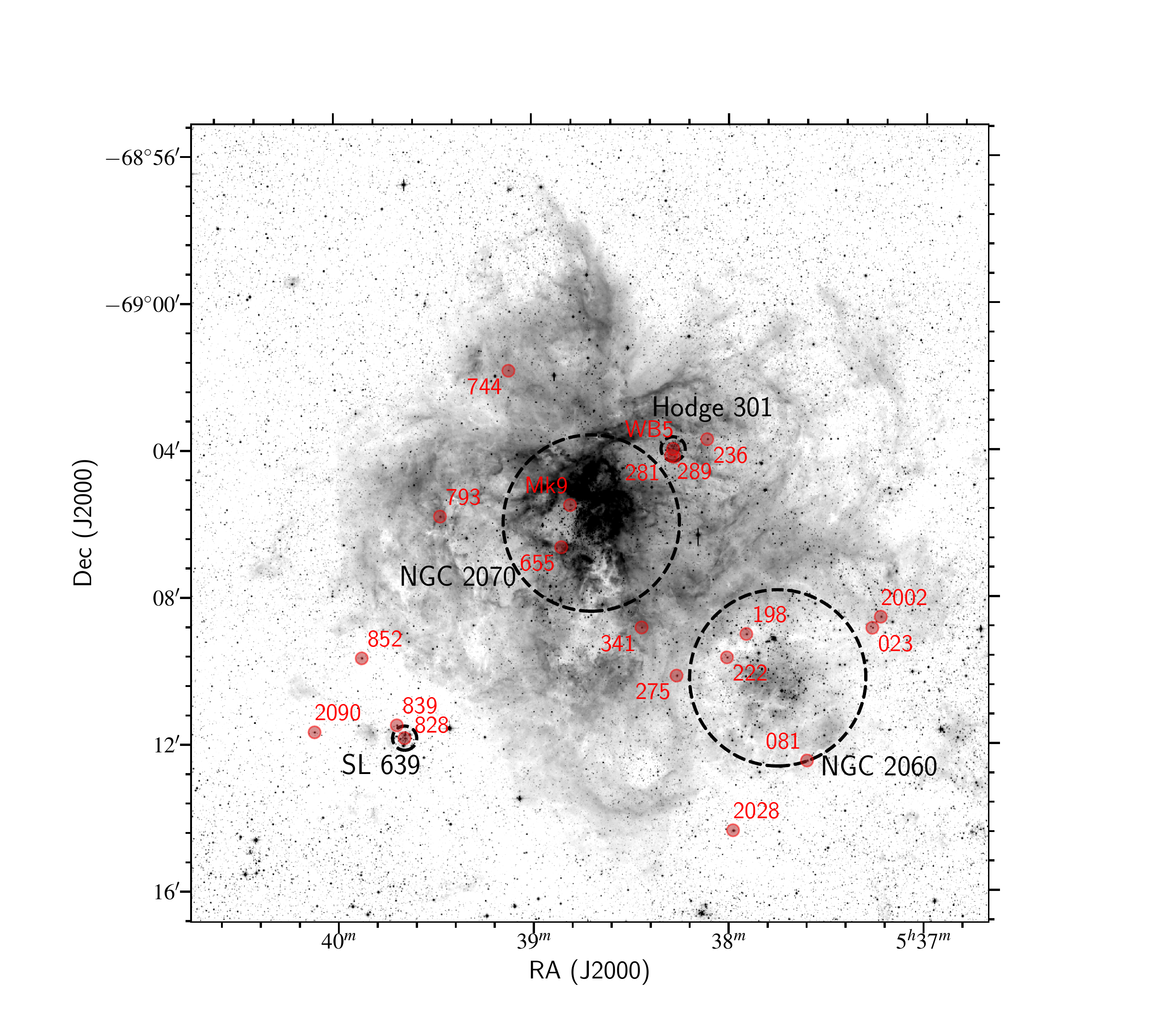}
\caption[]{Spatial distribution of the late-type stars in 30~Doradus and extent of the NGC\,2070, NGC\,2060, Hodge~301 and SL\,639 clusters \citep[as adopted by][in which the larger clusters have indicative diameters of $\sim$70\,pc and the smaller, older clusters have diameters of $\sim$9.5\,pc]{Evans_2015}. The image is from a $V$-band mosaic taken with ESO's Wide Field Imager on the 2.2\,m telescope at La Silla (under programme 076.C-0888).\label{spatial}}
\end{center}
\end{figure*}

The observational parameters of the resulting 20 candidate RSGs are summarised in Table~\ref{tb:parameters}. They can loosely be characterised as belonging to one of three groups: associated with the older clusters Hodge~301 and SL\,639, (loosely) associated with the young star-forming region NGC\,2060, and those in the field with no clear association. The locations of our sample in the 30~Dor region are shown in Fig.~\ref{spatial}, with the spatial extent of the four clusters as defined by \citet{Evans_2015} also shown.

Table~\ref{tb:parameters} includes the mean RVs for each target from cross-correlation of the LR02 spectra (3960--4564\,\AA) with a synthetic spectrum from a {\sc marcs} model atmosphere (see Patrick et al. submitted). All but one have RVs consistent with the systemic line-of-sight velocity of the 30~Dor region. The exception is VFTS\,793, with a significantly lower velocity of RV\,$=$\,187\,$\pm$\,1\,km~s$^{-1}$. The parallax ($p$) for VFTS\,793 from the {\em Gaia} Data Release 2 (DR2) catalog \citep[Gaia collaboration,][]{Lindegren_gaia2018} is $p$\,$=$\,0.1874\,$\pm$\,0.016\,mas, giving a distance modulus of 13.63\,mag ($\approx$5.3 kpc). As such, we consider this object as a foreground giant and exclude it from our subsequent analysis.

The RV estimates for the members of each cluster are in good agreement, and help to reveal three further stars which are potentially associated with the clusters.
VFTS\,236 is at a projected distance of only 14\,pc from Hodge\,301, and its RV estimate is near-identical to those for VFTS\,281 and 289. Similarly, the estimates for VFTS\,852 and 2090 (at projected distances of $\sim$36\,pc) are in good agreement with those for the two members of SL\,639, see Patrick et al. (subm.) for statistical arguments regarding membership of the respective clusters. For comparison, the mean RV for the remaining ten stars in Table~\ref{tb:parameters} (excl. VFTS\,793) is 271\,$\pm$\,15\,\kms, i.e. the three spatially-outlying stars from the clusters appear kinematically associated with them cf. the general velocity dispersion of the cool stars across the region. 
The radius adopted for these two clusters by \citet{Evans_2015} was a (knowingly conservative) {\em ad hoc} assumption of 20$''$ to delineate the sample to investigate the
cluster ages. That we find potentially associated stars at larger radii is not unexpected, and is analogous to the RSG population in the nearby NGC\,2100 cluster, which extends out to radii of nearly 30\,pc with a similarly small velocity dispersion \citep{p2100}. Thus, we consider these three stars (VFTS\,236, 852, 2090) as candidate members of their respective clusters.

Our analysis used magnitudes from the following sources: $V$-band from Paper~I, $I$-band from DENIS \citep{denis} and $JHK_{\rm s}$-bands from 2MASS \citep{2mass}. Significant spectral variability is thought to occur in only a relatively small fraction of RSGs in the LMC \citep[see discussion by][]{Bonanos_2009}. However, semi-regular photometric variability is seen in many RSGs and to account for the fact that we are using heterogeneous catalogues we adopted systematic photometric uncertainties of 0.2\,mag in the $V$-band and 0.1\,mag in the near-IR bands \citep[based on the average variability of well-studied RSGs,
e.g.][]{Josselin_2000,Kiss_2006,Yang_2011,Yang_2012}. We caution that a small fraction
of RSGs can undergo long-term variations of a magnitude or more in the visible, which may influence some of our results based on $V$-band magnitudes (cf. the near-IR), but our 
near-IR analysis should be robust to such effects.

\section{Determination of physical parameters}\label{analysis}

\subsection{Effective temperatures \& luminosities of red supergiants}

The effective temperatures of RSGs have been subject to a number of substantial studies in recent years. \citet{Levesque05} used optical spectrophotometry and revised {\sc marcs} models to arrive at an effective temperature scale that was approximately 10\% warmer than previously published values. They used this result to produce a `$V-K$' calibration of the effective temperature scale of RSGs and applied this to stars in the Magellanic Clouds \citep{Levesque2006,Levesque_2007}. However, \cite{Davies_2013} argued that such optical analyses are strongly influenced by the strong TiO bands in RSG spectra, which are thought to be formed further out in the atmosphere, yielding lower temperatures than spectral fits to the optical and near-IR continuum. 
The `$V-K$' method shows good agreement with theory in the derived temperatures for bright field RSGs in the Magellanic Clouds \citep{Levesque2006,Levesque_2007}. However, in the case of the 30 Dor sample that have substantially higher extinctions (cf. Table~\ref{tb:parameters}) the application of this method requires a precise extinction determination for each target. Clearly, the `$V-K$' calibration will be more sensitive to uncertainties in the $V$-band extinction, than to uncertainties in the $K_{\rm s}$-band.

Near-IR photometry can be useful to identify RSGs for spectroscopic follow-up \citep[e.g.][]{Patrick15}, for example, via their $J-K$ colours \citep{Nikolaev_2000}. However, it has not generally been used on its own to estimate physical parameters. Motivated by the above complications in using $V-K$, we investigated the use of near-IR photometry to estimate temperatures for our sample.
These two approaches are now discussed below, as well as the use of single-band photometry to estimate stellar luminosities independently of T$_{\rm eff}$, as advocated by \citet{Davies_2013}.

\begin{table*}
{\tiny
\caption{Observational properties of the late-type sample from the VLT-FLAMES Tarantula Survey (VFTS).}              
\label{tb:parameters}      
\centering                                      
\begin{tabular}{lc cc cc  cc ccl}          
\hline\hline                        
VFTS ID	& $\alpha$           & $\delta$	   &               $V$	 &$I$	 & $J$       & 	$H$   &  $K_{\rm s}$     &    A$_{V}$   & RV (LR02)     & Notes\\	
         &       (J2000)     &  (J2000)           &     (mag)                   &  (mag)      &      (mag)     &   (mag)       &   (mag)    &    (mag, $r\,<\,$1\farcm5)         &   (km s$^{-1}$)  &                               \\
\hline
023	&05 37 16.08 &$-$69 08 52.86&		  15.63  & 13.44 &  11.814 & 10.814 & 10.484 & 2.42 $\pm$ 0.46 &267.3 $\pm$ 2.6 & \\
081	&05 37 35.99 &$-$69 12 29.93&		  13.71  & 11.69 &  10.279 & $\phantom{0}$9.363  & $\phantom{0}$9.067  & 1.96 $\pm$ 0.25 &283.2 $\pm$ 2.4 & K4 (GF15)\\
198	&05 37 54.64 &$-$69 09 03.36&		  14.02  & 11.82 &  10.301 & $\phantom{0}$9.391  & $\phantom{0}$9.067  & 1.67 $\pm$ 0.52 &255.0 $\pm$ 1.4 & K4.5 Iab (GF15)\\
222	&05 38 00.57 &$-$69 09 41.64&		  14.86  & 12.93 &  11.456 & 10.588 & 10.312 & 1.40 $\pm$ 0.33 &250.4 $\pm$ 1.0 &\\
236	&05 38 06.61 &$-$69 03 45.25&		  13.80  & 11.90 &  10.536 & $\phantom{0}$9.677  & $\phantom{0}$9.413  & 1.14 $\pm$ 0.25 &258.5 $\pm$ 0.9 & Hodge 301 candidate \\
275	&05 38 16.00 &$-$69 10 11.39&		  14.22  & 11.47 &  $\phantom{0}$9.308  & $\phantom{0}$8.336  & $\phantom{0}$7.867  & 1.28 $\pm$ 0.22 &286.7 $\pm$ 1.0& M4 (GF15), M1.5 (L07)\\
281	&05 38 16.68 &$-$69 04 14.09&		  13.99  & 11.38 &  $\phantom{0}$9.786  & $\phantom{0}$8.906  & $\phantom{0}$8.552  & 1.14 $\pm$ 0.25 &260.3 $\pm$ 1.6& Hodge 301 member\\
289	&05 38 17.64 &$-$69 04 12.03&		  13.96  & 12.13 &  10.848 & 10.011 & $\phantom{0}$9.722  & 1.14 $\pm$ 0.25 &258.1 $\pm$ 1.7& Hodge 301 member \\
341	&05 38 26.69 &$-$69 08 52.70&		  14.02  & 11.66 &  $\phantom{0}$9.782  & $\phantom{0}$8.754  & $\phantom{0}$8.340  & 2.18 $\pm$ 0.81 &278.7 $\pm$ 2.9& K5 (GF15)\\
655	&05 38 51.19 &$-$69 06 41.29&		  15.13  & 13.02 &  11.497 & 10.598 & 10.260 & 2.00 $\pm$ 0.55 &282.0 $\pm$ 0.9&\\
744	&05 39 07.13 &$-$69 01 52.77&		  15.63  & 13.23 &  11.688 & 10.668 & 10.363 & 1.92 $\pm$ 0.91 &248.2 $\pm$ 3.1&\\
793	&05 39 28.18 &$-$69 05 50.49&		  13.58  & 12.33 &  11.422 & 10.735 & 10.616 & 1.75 $\pm$ 0.30 &187.4 $\pm$ 1.0& Foreground, K I (GF15) \\
828	&05 39 39.41 &$-$69 11 52.05&		  14.52  & 11.60 &  $\phantom{0}$9.864  & $\phantom{0}$8.965  & $\phantom{0}$8.524  & 2.50 $\pm$ 0.30 &247.0 $\pm$ 1.1& SL 639 member, M (GF15)\\
839	&05 39 41.78 &$-$69 11 31.01&		  14.64  & 12.03 &  10.280 & $\phantom{0}$9.352  & $\phantom{0}$8.935  & 2.50 $\pm$ 0.30 &248.0 $\pm$ 1.2&SL 639 member\\
852	&05 39 52.39 &$-$69 09 41.26&		  14.30  & 12.21 &  10.682 & $\phantom{0}$9.733  & $\phantom{0}$9.407  & 2.15 $\pm$ 0.21 &243.2 $\pm$ 3.0 & SL 639 candidate \\
2002	&05 37 13.50 &$-$69 08 34.65&		  14.21  & 11.57 &  $\phantom{0}$9.720  & $\phantom{0}$8.721  & $\phantom{0}$8.292  & 2.46 $\pm$ 0.55 &284.3 $\pm$ 3.4& M3 (GF15) \\
2028	&05 37 58.67 &$-$69 14 24.07&		  13.27  & 11.06 &  $\phantom{0}$9.532  & $\phantom{0}$8.716  & $\phantom{0}$8.383  & 1.20 $\pm$ 0.33 &274.0 $\pm$ 2.5& G5 Ia (GF15) \\
2090	&05 40 07.01 &$-$69 11 41.50&		  14.88  & 12.65 &  11.035 & 10.067 & $\phantom{0}$9.717  & 2.50 $\pm$ 0.30 &247.1 $\pm$ 1.0& SL 639 candidate \\
WB97 5 &05 38 17.01 &$-$69 04 00.98 &    13.60  & 11.58 &  10.077 & $\phantom{0}$9.193  & $\phantom{0}$8.866  & 1.17 $\pm$ 0.24 & \ldots & Hodge 301 member \\
Mk 9      &05 38 48.48 &$-$69 05 32.58 &    13.62  & 10.86 &  9.173 & $\phantom{0}$8.297  & $\phantom{0}$7.869  & 1.92 $\pm$ 0.30 & \ldots & M3.5 Ia (GF15)\\
\hline                                             
\end{tabular}
\tablefoot{The final column indicates membership of Hodge~301 or SL\,639, and published spectral classifications from \citet[][GF15]{new_smclmc} and \citet[][L07]{Levesque_2007}. Radial velocity (RV) estimates are the averages from the VFTS observations with the LR02 setting.}}
\end{table*}

\subsubsection{V -- K method}\label{VK}
We initially estimated stellar parameters using the $V$\,$-$\,$K$ method from \citet{Levesque05}. Aside from the issues arising from using the TiO bands, we were interested to investigate this approach for comparison with other methods. As discussed, many of our sources have high extinction, so the most critical aspect with this method is to obtain reliable estimates for the optical extinction ($A_{V}$) to each target.
For this purpose we used the mean reddening of (morphologically normal) O-type stars from \citet{Walborn_2014} within a search radius of 1\farcm5 of each candidate RSG (typically yielding 3--5 O-type stars per target). Given estimates of $E(B-V)$ for each target and adopting a ratio of total-to-selective extinction of $R_{V}$\,$=$\,4.48\,$\pm$\,0.24 \citep{http}, we estimated the line-of-sight extinction from $A_{V}$\,$=$\,$E(B-V)$\,$\times$\,$R_{V}$. The resulting extinction estimates and their uncertainties are listed in Table~\ref{tb:parameters}, where the values for most of the stars in Hodge~301 and SL\,639 are identical simply because they are located so close to each other. Our estimated reddening toward Hodge~301 is in good agreement with the mean value of $E(B-V)$\,$=$\,0.28\,$\pm$\,0.05 from \citet{Grebel_2000}.

We then used the calibrations of T$_{\rm eff}$ and bolometric corrections (BC$_{K}$) as a function of $V-K$ colour from \citet{Levesque2006,Levesque_2007}. Their technique was based on fits to spectrophotometric observations of 36 RSGs in the LMC with synthetic spectra calculated from {\sc marcs} atmosphere models \citep{Gustafsson_2003,Gustafsson}. The general precision of the method, i.e. the standard deviation of temperature differences between the {\sc marcs} model fitting of the TiO region and the $V$\,$-$\,$K$ calibration is $\sim$100\,K \citep[see][]{emily_book}.  

To apply this technique to our sample we dereddened our sources using: $(V-K_{\rm s})_{0}$\,$=$\,$(V-K_{\rm s}$)\,$-$\,0.87\,$A_{V}$, based on $A_{K}$\,$=$\,0.13\,$A_{V}$ \citep[from the $R_{V}$ dependent extinction law of][adoping $R_{V}$\,$=$\,4.48]{Donnell_1994}, then calculated bolometric magnitudes
from: $M_{\rm Bol}$\,$=$\,$K$\,$-$\,0.13\,$A_{V}$\,$-$\,DM\,$+$\,BC$_{K}$,
in which we adopt a distance modulus to the LMC of 18.5\,mag. Estimates of T$_{\rm eff}^{(V-K)}$ and L$^{(V-K)}$ using their calibration are given in Table~\ref{tb:derived}.

\subsubsection{J -- K method}
\citet{JK_corr} presented bolometric corrections for late-type stars in the Clouds based on the $(J-K_{\rm s})$ colours for 90 stars. \citet{Dorda_2016} employed this relation for RSG stars, entailing transforming the photometric system\footnote{$(J-K_{\rm s})_{\rm AAO}=[(J-K_{\rm s})_{\rm 2MASS}\,-\,0.013$]/0.953} and then dereddening the colours for each target. They then used the results from \citet{JK_corr} to estimate BC$_{K}$, thence luminosity (L$^{(J-K)}$), analagous to the method in Section~\ref{VK}. We also investigate this approach for our sample.

\begin{figure*}
\begin{center}
\includegraphics[width=19cm]{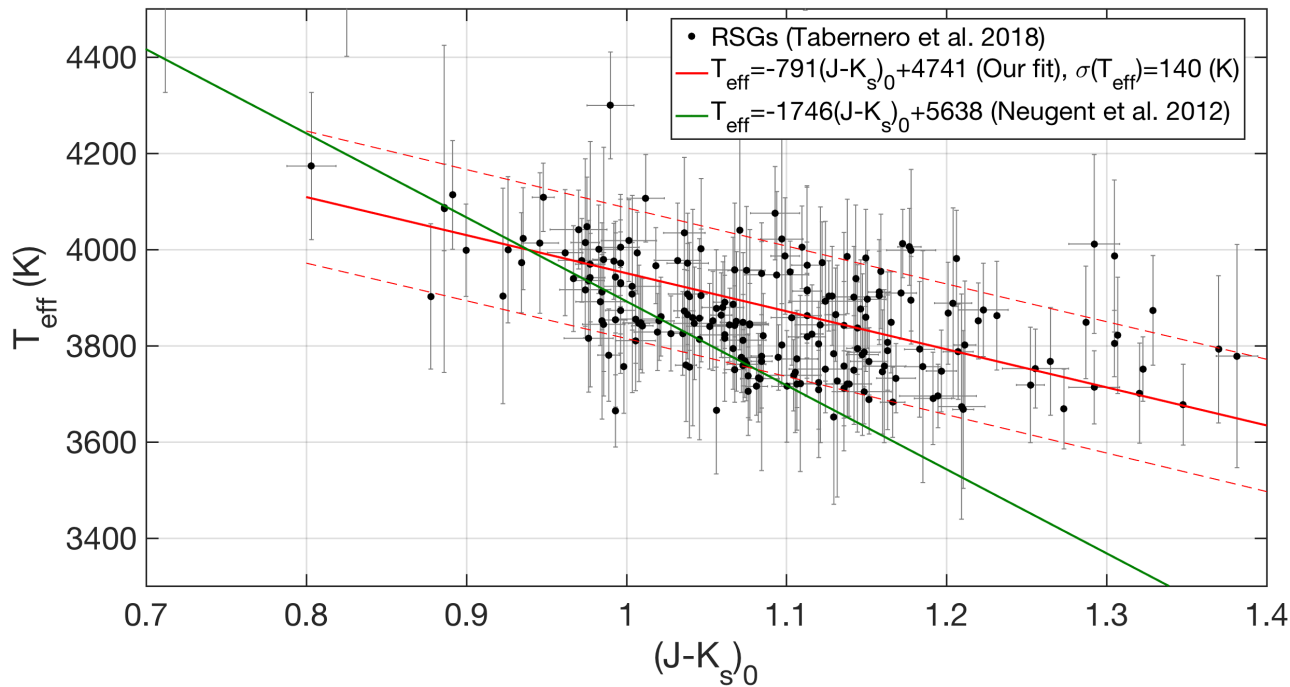}
\end{center}
\caption[]{Effective temperature (T$_{\rm eff}$) vs. $(J-K_{\rm s})_{0}$ colour for the 217
RSGs in the LMC from \citet{Tabernero_2018}. The red line is our linear fit to their data and the green line shows a previous calibration from a sample of yellow SGs and RSGs by \cite{2012_Neugent}.}
\label{T_JK}
\end{figure*}

The $(J-K_{\rm s})$ colours also appear to be useful to estimate effective temperatures of RSGs. \citet{Tabernero_2018} presented temperatures for 217 RSGs in the LMC from fits to
spectra of the 8400--8800\,\AA\/ region using synthetic spectra calculated from 1D Local Thermodynamic Equilibrium (LTE) {\sc kurucz} models \citep{Kurucz_2012}.  This region is relatively free of telluric
and molecular bands, and includes several Fe\,{\scriptsize I} lines as well as the strong Calcium Triplet lines in RSGs, e.g. Fig.~2 from \citeauthor{Tabernero_2018} \citep[see also, e.g.][]{britavskiy14}. As shown in Fig.~\ref{T_JK}, with the benefit of the large sample of results from \citeauthor{Tabernero_2018}, a good correlation between T$_{\rm eff}$ and $(J-K_{\rm s})$ is revealed, with a linear fit (valid for 0.8 mag \,$<$\,$(J-K_{\rm s})$\,$<$\,1.4 mag) described by: 
\begin{equation}\label{Teff_relation} 
{\rm T}_{\rm eff}^{(J-K)} = -791 \times (J-K_{\rm s})_{0} + 4741, 
\end{equation} 
with a standard deviation of $\sigma$(T$_{\rm eff})$\,$=$\,140\,K. To use this relation it is necessary to deredden the colour, i.e. $(J-K_{\rm s})_{0}$\,$=$\,$(J-K_{\rm s})$\,$-$\,$E(J-K)$. From \citet{Schlegel_1998} the near-IR reddening can be derived as $E(J-K)$\,$=$\,0.535$E(B-V)$, assuming $R_{V}$\,$=$\,3.1, with the reddening value of $E(B-V)$ specific to each target. A calibration of T$_{\rm eff}$ vs. $(J-K_{\rm s})_{0}$ from \citet{2012_Neugent} is also
shown in Fig.~\ref{T_JK}, which was derived using a sample of yellow SGs 
($(J-K_{\rm s})$\,$<$\,0.9 mag) and candidate RSGs in the LMC. Their hotter sample and
their fits with varying log~$g$ values (cf. the constant value of zero adopted by
\citeauthor{Tabernero_2018}) both contribute to the difference in slopes.

Our estimates of temperature (T$_{\rm eff}^{(J-K)}$, from Eqn.~\ref{Teff_relation}) and luminosity (L$^{(J-K)}$) from this method are given in Table~\ref{tb:derived}, in which the uncertainties on the former are the dispersion of 140\,K in the calibration combined with the uncertainty in T$_{\rm eff}$ arising from the uncertainty on the extinction. For consistency, the latter includes use of the same $E(J-K)$ relation as for the \citeauthor{Tabernero_2018} sample (although using $R_V$\,$=$\,4.48 would lead to only a small change in colour, with minimal impact on T$_{\rm eff}$ cf. the spread in Fig.~\ref{T_JK}).

\subsubsection{Single-band photometry}
Empirical luminosity calibrations for single-band photometry of RSGs in the LMC and SMC were given by \citet[][their Table~4]{Davies_2013}. These calibrations assumed that the bolometric correction for the RSGs is constant for each given photometric band, a consequence of the near-uniform temperatures (to $\pm$100\,K) of the stars analysed by \citeauthor{Davies_2013}
To investigate this method for our sample we used the available $I$-, $J$- and $K_{\rm s}$-band magnitudes, allowing us to test for the effects of extinction and to understand which band gives the most robust results. For each band ($x$), we calculated the absolute magnitude for each target as: $M_{x}$\,$=$\,$m_{x}$\,$-$\,$A_{x}$\,$-$\,DM, with estimates of extinction of: $A_{I}$\,$=$\,0.54\,$A_{V}$, $A_{J}$\,$=$\,0.32\,$A_{V}$, $A_{K}$\,$=$\,0.13\,$A_{V}$, based on the adopted value of $R_{V}$=4.48 by applying the extinction law of \citet{Donnell_1994}. While the extinction coefficient for each band depends on spectral type \citep{vanLoon_2003}, taking into account that all our targets have K or early M spectral types, the resulting difference in extinction is very small and we did not take it into account.

The resulting luminosity estimates (L$^{(I)}$, L$^{(J)}$, and L$^{(K)}$) are listed in Table~\ref{tb:derived}. The internal dispersion of the three bands for our stars was $\sigma$(log(L/L$_{\sun}$))\,$=$\,0.05, although we note the calibration itself was limited to a sample of 19 RSGs in the Clouds \citep{Davies_2013}. 
As pointed out by \citet{Davies_2013} a major advantage of this approach is that the resulting stellar luminosities are relatively insensitive to errors in the effective temperature since we are near the flux maximum of the spectral energy distribution. The resulting uncertainties in the luminosities for each method listed in Table~\ref{tb:derived} consist of the internal dispersion of the methods together with the individual uncertainties on the $A_{V}$ and $R_{V}$ values.


\begin{figure}
\begin{center}
\includegraphics[width=9.5cm]{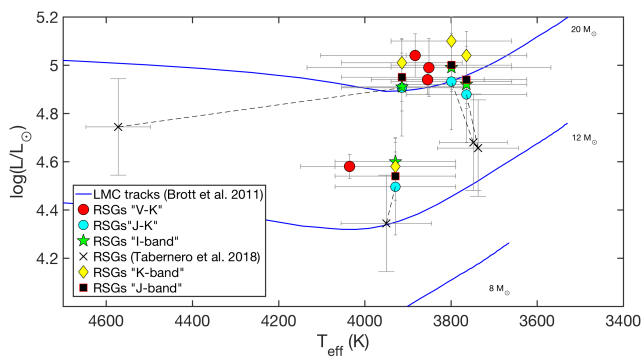}
\end{center}
\caption{Hertzsprung--Russell diagram showing results for four stars from the photometric
methods discussed in Section~\ref{analysis} compared with published results
from \citet[][linked by the dashed lines to our results]{Tabernero_2018}.}
\label{alicante}
\end{figure}

\begin{figure}
\begin{center}
\resizebox{\hsize}{!}{\includegraphics{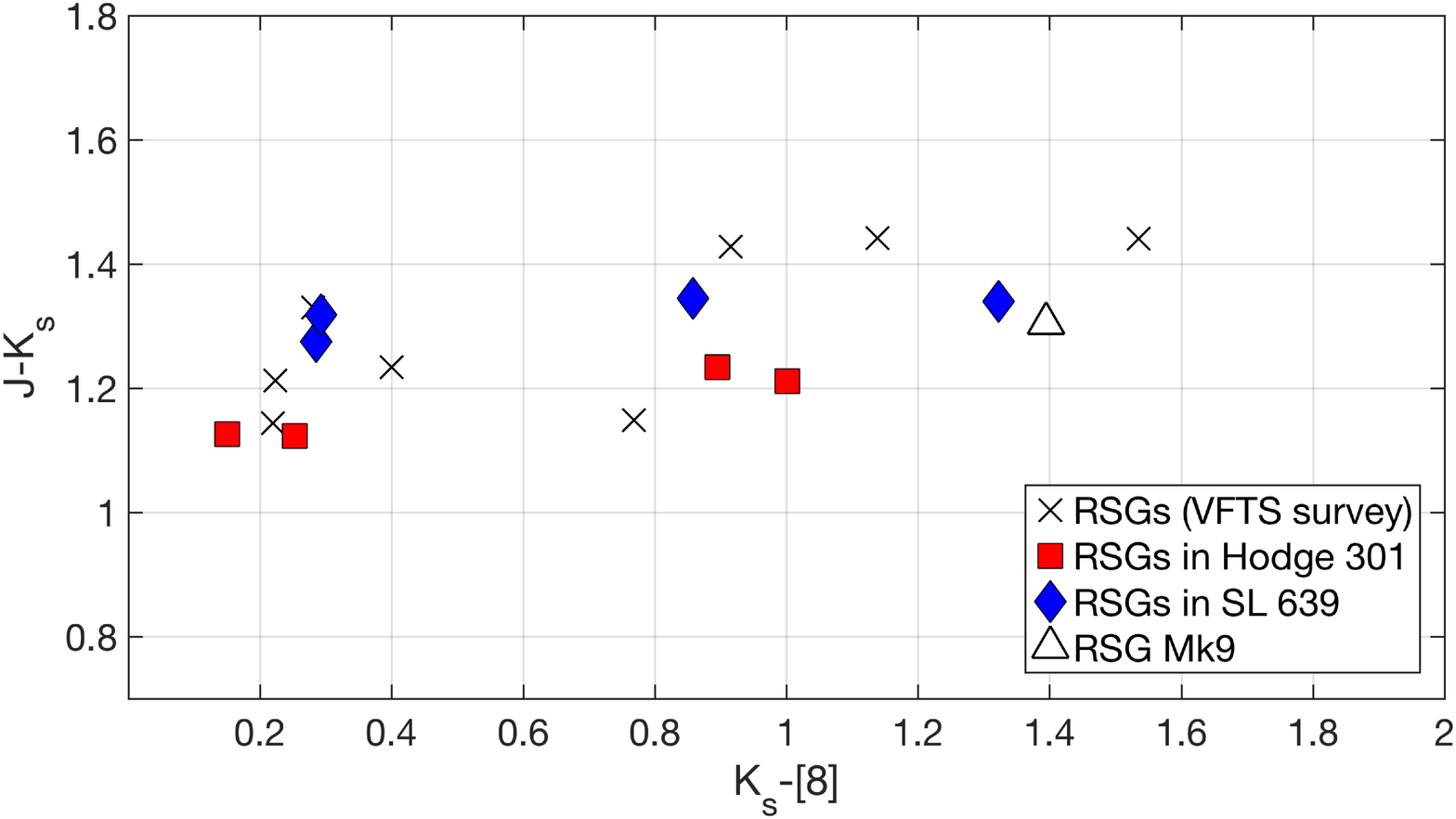}}
\resizebox{\hsize}{!}{\includegraphics{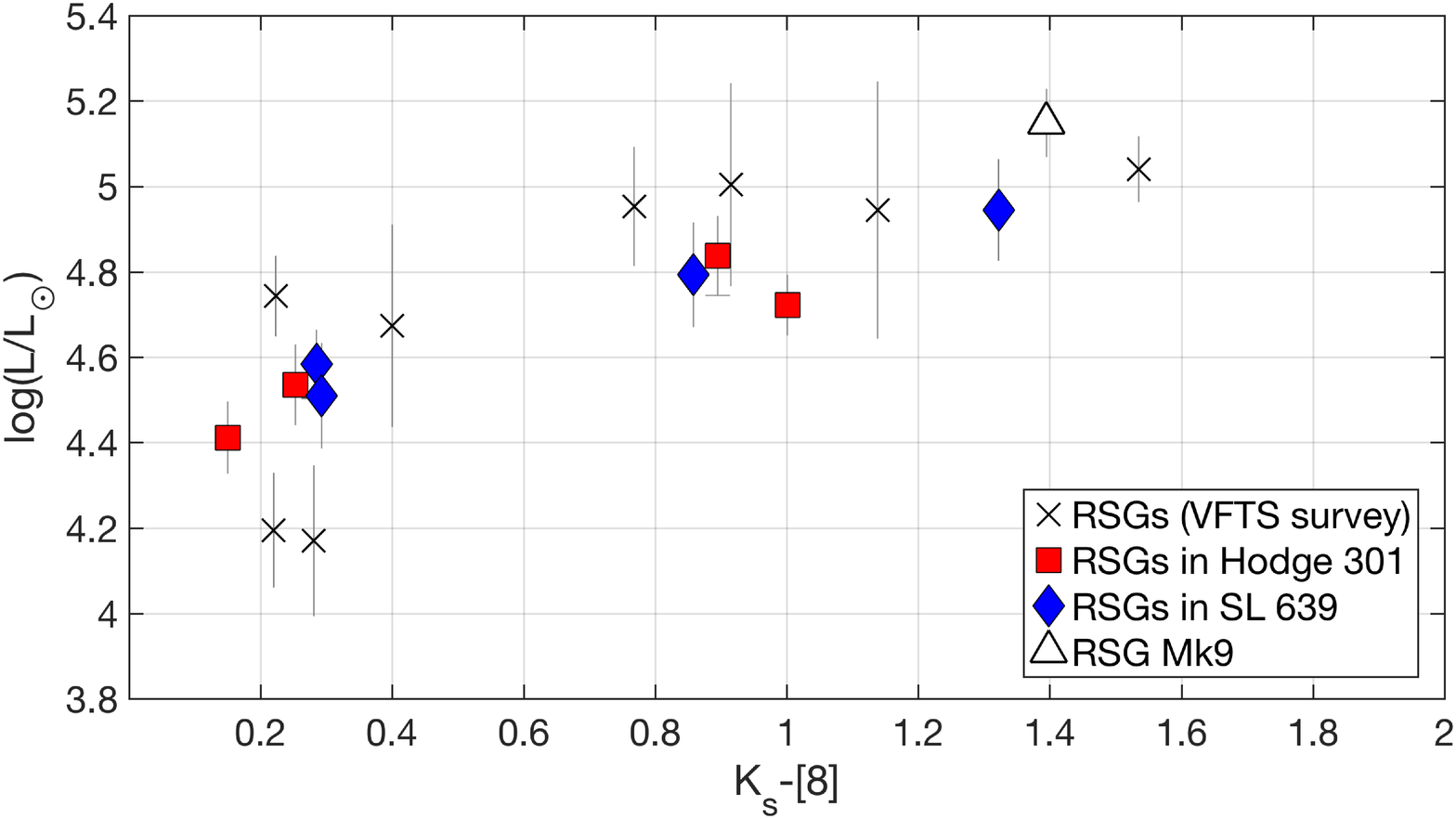}}
\end{center}
\caption{{\it Upper panel:} $(J-K_{\rm s})$ vs. $(K_{\rm s}-[8.0])$ diagram for our sample. {\it Lower panel:} Luminosity vs. $(K_{\rm s}-[8.0])$ for the same stars.}
\label{mid_color}
\end{figure}

\begin{figure*}
\begin{center}
\resizebox{\hsize}{!}
{\includegraphics{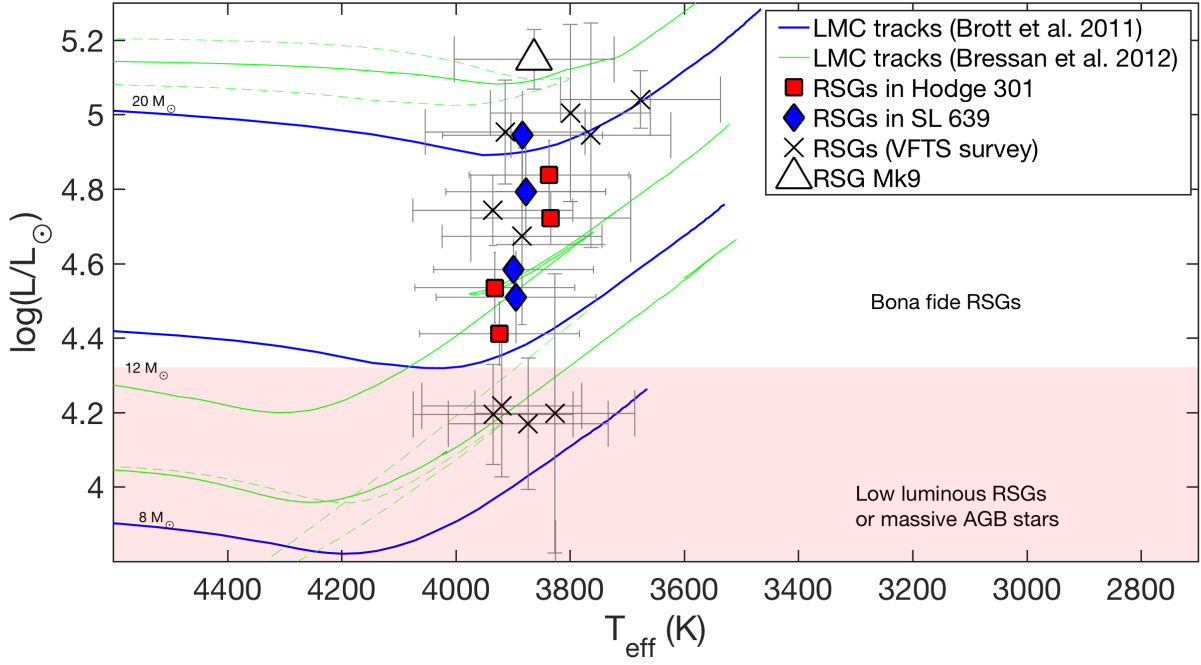}}
\end{center}
\caption[]{Hertzsprung--Russell diagram for our targets, in which temperatures were estimated using the $J$\,$-$\,$K$ approach and luminosities estimated from $J$-band magnitudes. The  classifications of the targets based on our analysis are highlighted by the comments and filled region. Blue loops in the {\sc parsec} evolutionary models from \citet{Bressan_2012} are marked by dashed lines (see text for details).}
\label{HR_all}
\end{figure*}

\begin{figure*}
\begin{center}
\resizebox{\hsize}{!}
{\includegraphics{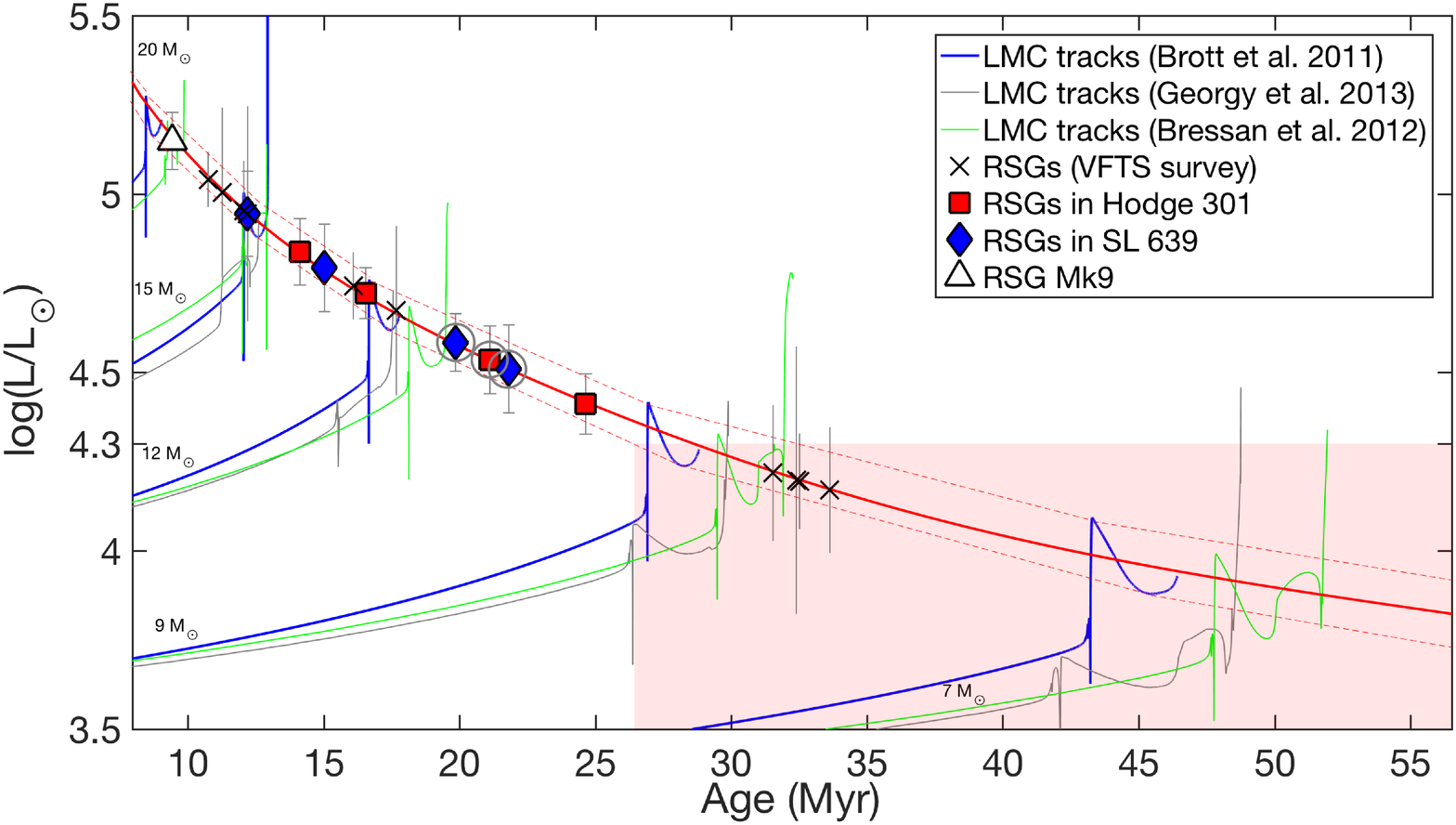}}
\end{center}
\caption{Luminosity--age diagram for our sample of RSGs compared with evolutionary tracks for LMC metallicity from \citet{Brott2011}, \citet{Georgy_2013} and \citet{Bressan_2012}. The luminosities that RSGs are expected to occupy during the He-burning phase from the \citet{Brott2011} models are shown by the red dashed lines. The red solid line corresponds to a fourth-degree polynomial interpolation of the RSG region: Age\,$=$\,$-$0.413L$^{4}$$-$3.868L$^{3}$+128.081L$^{2}$$-$796.823L+1534.899 (Myr), where L is in units of log(L/L$_{\sun}$). This interpolation can be used for ages of 8 to 55\,Myr. The filled red area highlights ages that would be excluded by a luminosity threshold of log(L/L$_{\sun}$) = 4.3 dex being adopted as a lowest RSG luminosity limit according to Fig.\ref{HR_all}. Using our derived stellar luminosities we can place each star in the RSG region to read off an evolutionary age, or age range, as shown. The cluster candidates are marked by grey circles.}
\label{vfts_lage_all}
\end{figure*}

\begin{figure}
\begin{center}
\resizebox{\hsize}{!}
{\includegraphics{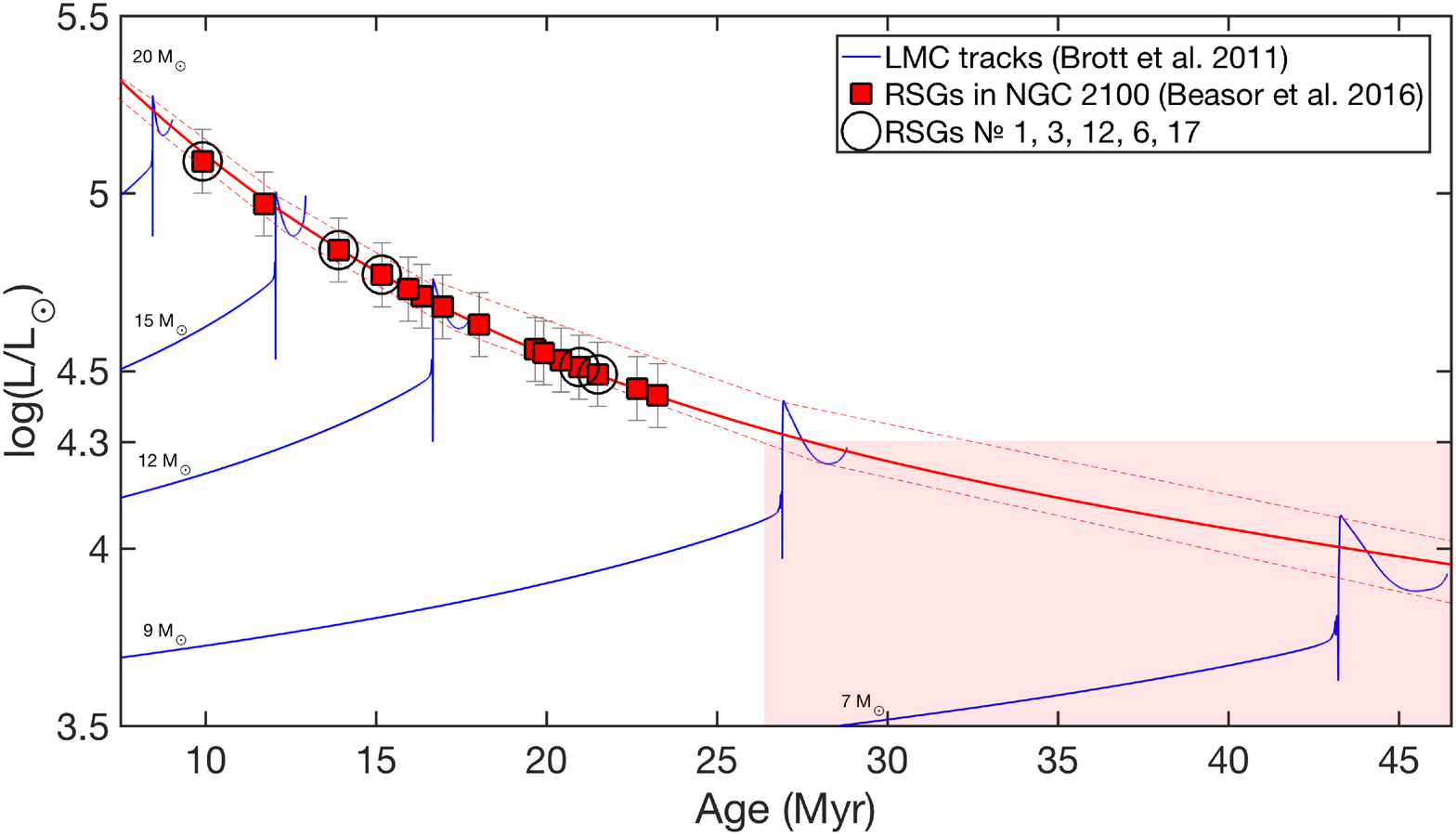}}
\end{center}
\caption{Luminosity--age diagram as in Fig.~\ref{vfts_lage_all} but using results from \citet{Beasor_2016} for RSGs in the relatively nearby NGC 2100 cluster. The five most distant
RSGs from the visual cluster centre are highlighted to investigate potential spatial effects
(see text for discussion).}
\label{vfts_lage_ngc2100}
\end{figure}

\subsubsection{Comparison of the three methods}

Estimates of temperature and luminosity for four of our targets are available from \citet{Tabernero_2018}. Our results from the three methods outlined above are compared
with theirs in the Hertzsprung--Russell (H--R) diagram in Fig.~\ref{alicante}. \citeauthor{Tabernero_2018} calculated luminosities using the same J--K method as here, but without correcting for extinction, hence our estimates are all larger than theirs. Note that the three methods explored here give similar luminosity estimates once extinction is taken into account. 
Indeed, for our heavily reddened sample, it is critical to correctly account for interstellar extinction to obtain reliable estimates of their luminosities.

The star with a large difference in T$_{\rm eff}$ cf. \citeauthor{Tabernero_2018} is
VFTS\,2028 (T$_{\rm eff}^{(J-K)}$\,$=$\,3914\,$\pm$\,143\,K here, cf. the published value of 4572\,$\pm$\,75\,K). From inspection of Fig.~\ref{T_JK} we note
that the published value is a substantial outlier cf. the overall population of RSGs. The reason for such an outlier is that the T$_{\rm eff}$ vs. (J-K$_{\rm s})_{0}$ calibration is showing the average trend, without taking into account information on the luminosity class and spectral type of RSGs. As shown by \citeauthor{Tabernero_2018}, at a given bolometric luminosity the temperatures of RSGs can vary with different spectral types and luminosity classes (Ia--Ib), such that the temperamperature difference may be partly explained by their assumption of log\,$g$\,$=$\,0. Published estimates of T$_{\rm eff}$ and L are also available for VFTS\,275 from \citet[][their LMC170452]{Levesque_2007}. This object has some photometric and spectral type variability (as reported by \citeauthor{Levesque_2007}), but our estimates are in reasonable agreement.


Based on the above discussion we therefore adopt L$^{(J)}$ and T$_{\rm eff}^{(J-K)}$ in our subsequent analysis. The locations of our RSGs in the H--R diagram are shown in Fig.~\ref{HR_all} together with appropriate evolutionary tracks from \cite{Brott2011} and \citet[][{\sc parsec} models]{Bressan_2012}.
The selection of L$^{(J)}$ is to mitigate against IR excesses arising from strong mass-loss, which start to be significant in the $H$- or $K$-bands. Moreover, the peak of the spectral energy distribution for RSGs occurs in the $I$- and $J$-bands, and the effects of extinction in the $J$-band are relatively low compared to the optical. Indeed, luminosities estimated from integration of the spectral energy distributions for RSGs in dwarf irregular galaxies in the Local Group, using either the $I$- or $J$-band, are found to be in good agreement (Britavskiy et al. in prep.). 

We note that RSGs may experience strong mass loss that would lead to significant circumstellar extinction, as seen in some of the most luminous RSGs \citep[e.g.][]{Massey2005, Davies_2008, Beasor_2018}. Fig.~\ref{mid_color} shows the $(J-K_{\rm s})$ vs. $(K_{\rm s}-[8.0])$ colour-colour diagram for our sample, together with the distribution of L$^{(J)}$ vs. $(K_{\rm s}-[8.0])$, where the 8.0\,$\mu$m magnitudes ([8.0]) are from the {\it Spitzer Space Telescope} Legacy Survey \citep[SAGE,][]{Meixner_2006}. Note that three of our targets are not included in Fig.~\ref{mid_color} as there were no [8.0] magnitudes available. A significant mid-IR excess \citep[$K_{\rm s}$\,$-$\,{[}8.0{]}\,$>$\,0.5\,mag;][]{vanLoon_2003} is present in some of the more luminous RSGs (lower panel of Fig.~\ref{mid_color}). This indicates the presence of a dusty circumstellar envelope around these RSGs, and consequently, our luminosity estimates for these RSGs may be slightly underestimated. However, for the less luminous stars, we do not see similar evidence for circumstellar dust emission. 


\subsection{Determination of masses, radii and ages}

Stellar radii for our sample were derived from the Stefan--Boltzmann relation, i.e.: R/R$_{\odot}$\,$=$\,(L$^{(J)}$/L$_{\odot}$)$^{0.5}$(T$_{\rm eff}^{(J-K)}$/5770)$^{-2}$. The uncertainties on the radii were calculated with the help of Monte Carlo simulations by using the standard deviations in the T$_{\rm eff}$ and L estimates. To estimate the initial masses of each target we interpolated their position in the H--R diagram, compared with the evolutionary tracks for an initial rotation rate of 150\,km~s$^{-1}$ from \citet[][which gives a detailed discussion of the treatment of rotation]{Brott2011}. The interpolation was based on the tracks for 7, 9, 12, 15, 20, 25 M$_{\odot}$, to which we fit the position of RSGs based on their T$_{\rm eff}$ and log(L/L$_{\odot}$). The uncertainties of the derived masses were based on uncertainties of effective temperature and luminosity for each RSG. Obviously, the uncertainties in the luminosities of the targets play a major role in the total error budget of their mass. We then estimated logarithmic gravities from log~$g$\,$=$\,log(M/M$_{\odot}$)\,$-$\,2log(R/R$_{\odot}$)\,$+$\,4.438. The resulting values of radius, mass and gravity for each target are listed in Table~\ref{tb:derived}.

To investigate the ages of our targets we used a luminosity--age (L--Age) diagram, as
shown in Fig.~\ref{vfts_lage_all}. This includes evolutionary tracks of single stars for the LMC with an initial rotational velocity of 150\,km\,s$^{-1}$ from \citet{Brott2011}. The advantage of this approach is that it is not dependent on the estimated temperatures and the expected region for RSGs (highlighted by the dashed red lines) arises from steep increases in luminosity (from the start of the He-burning phase to the Hayashi limit at the end). 
From the tracks in Fig~\ref{vfts_lage_all} it is clear that the He-burning phase begins earlier than the highlighted range, when the luminosity begins to significantly rise. However, to observe such stars during this short initial rise ($\approx$60\,000 years) is very unlikely compared to the total duration of the He-burning phase ($\sim$1.9 Myr). Therefore we do not consider this early stage further. 

To place our targets in Fig.~\ref{vfts_lage_all} we matched the observational luminosities L$^{(J)}$ to the expected region for RSGs, yielding age estimates. The uncertainties in luminosity were then used to quantify the uncertainties on the age determination, simply by fitting the min/max L$^{(J)}$ within the RSG region in the figure. Estimated ages for each target are presented in Table~\ref{tb:derived}.

To assess the impact of using different evolutionary models in the L--Age diagram, Fig.~\ref{vfts_lage_all} also includes the `Geneva' tracks from \citet{Georgy_2013} and the {\sc parsec} models of \citet{Bressan_2012}. As in the models from \citet{Brott2011}, the early onset of He-burning in the {\sc parsec} models leads to a rapid increase in luminosity, meaning that the majority of the RSG lifetime spans a similarly small luminosity range for given initial mass.

For low-mass RSGs (M\,<15\,M$_{\sun}$) blue loops appear in the Geneva tracks, and at lower masses (M\,$<$\,12\,M$_{\sun}$) such blue loops also appear in the {\sc parsec} models. This is due to the different treatment of mass loss and overshooting of the convective core in the models \citep{2014_Castro}. \citet{Brott2011} calibrated the overshooting parameter using $\approx$15\,M$_{\sun}$ stars, while the {\sc parsec} models were calibrated using stars more massive than 14\,M$_{\sun}$. In contrast, the Geneva models tuned this parameter for stars with M\,$<$\,8\,M$_{\sun}$. We therefore chose the \citet{Brott2011} and \citet{Bressan_2012} tracks as more applicable to our RSG sample (see Table~\ref{tb:derived}); the differences between the age estimates using these two sets of tracks are small (within 1\,Myr). This demonstrates that the uncertainties in the RSG luminosities play a more significant role in estimating their ages than employing different evolutionary tracks.



From the L--Age diagram we can see that the generally adopted minimum initial mass for RSGs of 8\,M$_{\odot}$ is equal to 35\,Myr. However, taking into account the luminosity range of the He-burning phase, the age limit should be extended down to $\sim$26\,Myr. The targets below this mass limit are generally considered as intermediate-mass stars, including massive AGB stars. There is no strong morphological separation by using only luminosity as a parameter to distinguish RSGs from AGB stars. However, we consider the four targets from our sample with 4.0\,$<$\,log(L/L$_{\sun}$)\,$<$\,4.3 as massive AGB stars or low-mass RSG candidates (see Fig.~\ref{HR_all} and Fig.~\ref{vfts_lage_all}). Based on the L--Age diagram, we consider targets as bonafide RSGs if they have log(L/L$_{\sun}$)\,$>$\,4.3.

With spectroscopy at $R$\,$\lesssim$\,20\,000 there is currently no reliable observational method to distinguish RSGs from massive AGBs \citep[e.g.][]{2005_vanloon,2010_Doherty,2017_Doherty}. 
Atomic lines from, e.g., lithium and/or rubidium can potentially be used to distinguish these types of stars spectroscopically. These elements are produced in massive (M\,$>$\,4\,M$_{\sun}$) O-rich AGB stars during the short, so-called `hot bottom burning' phase. Thus, an overabundance of these elements can be observed during some AGB phases \citep{2006_rb,2007_li}. For definitive line identifications to investigate this further we  require higher-resolution spectroscopy and over a wider wavelength range than the VFTS data.

Our analysis relies on the evolutionary models of \citet{Brott2011}, in which uncertain physics, in particular, rotational mixing and convective core overshooting, have been carefully calibrated to spectroscopic observations of massive stars in the LMC obtained within the VLT-FLAMES Survey of Massive Stars \citep{Evans_2005A&A...437..467E}. Nevertheless, there are appreciable remaining uncertainties in models of massive-star evolution \citep{Langer_2012} which lead to systematic differences in corresponding evolutionary tracks \citep[e.g.][]{Georgy_2013,Choi_2016,Limongi_2018}. An assessment of these systematic differences, which are not reflected in our derived error bars, is beyond our ability to assess at this time. That is, even though they reflect our best estimate, the absolute values of the derived ages and masses may still be subject to changes. However, the age and mass differences will be affected much less, such that our hypothesis of an age spread partly arising from red stragglers should still hold.

\section{Discussion}\label{sec:discussion}

Assuming our objects have evolved as single stars we can see from Fig.~\ref{vfts_lage_all} that our sample of bona fide RSGs have an age range of approximately 9--24\,Myr. This is not surprising since, despite the young ages of NGC\,2060 and NGC\,2070 \citep[e.g.][]{2017A&A...600A..81R,Schneider_2018} there is also a significant population of massive stars with similar ages throughout the 30~Dor region \citep[e.g.][]{Sabbi_2016,Schneider_2018}. Indeed, seven of our RSG sample are associated with either Hodge~301 or SL\,639. \citet{Evans_2015} derived ages of 10--15\,Myr for both clusters from examination of the properties of the massive stars near their main sequence turn-offs. This `young' age for Hodge~301 contrasts strongly with the estimate of 26.5 to 31.5\,Myr from \citet{Cignoni} from analysis of pre-main-sequence turn-on stars in the observed colour--magnitude diagram. \citet{Cignoni} discussed potential reasons for this difference, noting that the ages implied by the blue and red supergiants in this cluster also differ from their turn-on age, being intermediate to their turn-off and turn-on ages. In the present work we find that the RSGs associated with Hodge~301 have ages with a significant age spread of 14--24\,Myr and, similarly, the RSGs in SL\,639 have an age range of 12--22\,Myr.

Binary evolution might help understand the large dispersion in the derived cluster ages.  Similar to the ubiquitous blue straggler phenomenon, i.e. main sequence stars found above the main sequence turnoff \citep{Schneider_2016}, binary evolution can also produce red supergiants above the red giant branch of the single stars in a star cluster, i.e. red stragglers. To illustrate this, we consider our results for Hodge 301, where the lowest luminosity RSG (VFTS 0289) has an estimated age of 24 Myr, corresponding to the lifetime of a single star of ~10 $M_{\odot}$.  The merger of two initially 7\,$M_{\odot}$ stars at an age of 24 Myr would produce a ~14 $M_{\odot}$ star, which would soon thereafter evolve into a RSG. Its luminosity, if interpreted with only single stars in mind, would lead to an age of the cluster comparable to the lifetime of a 14 $M_{\odot}$ star, i.e., only about 15 Myr. Therefore, the derived cluster age would be $\sim$ 60\% too small.

The example of two merging 7 $M_{\odot}$ stars, though not unrealistic, was chosen to demonstrate the possibility of producing red stragglers, and to show the order of magnitude of the effect on the derived cluster age. Of course, binaries with any initial mass ratio may produce mergers \citep[see Fig.\,12 of][]{2001_Wellstein}. Binary population synthesis calculations are required to derive more accurate predictions of the red straggler distribution in star clusters \citep[unfortunately unavailable in the recent work by][]{Dorn_2018}. However, as red stragglers are merely evolved blue stragglers, their fraction among the red supergiants in a well populated cluster can be expected to be significant, given that the blue straggler fraction near the main sequence turnoff is found to be up to 30\% in young open clusters \citep{2015_Schneider}.

We note here that the latter effect, of blue stragglers, was not taken into account in the age determination of the turn-off stars in these clusters by \citet{Evans_2015}. Indeed in a recent study of an analagous Galactic cluster, NGC\,3293, \citet{Proffitt_2016} suggested that its brightest and apparently youngest blue supergiants might indeed be blue stragglers that are the result of binary evolution. The apparent age spread of the turn-off stars in Hodge\,301 and SL\,639 is further complicated as known Be stars are included in the H--R diagram presented by \citet{Evans_2015}. 
As well as being intrinsically variable, their stellar parameters are highly uncertain due to the impact of the circumstellar disk on extinction, apparent magnitude, and veiling of their absorption lines by the disk continuum emission \citep{Lennon_2005,Dunstall_2011}. 
Both clusters have significant numbers of Be stars; four of the 15 blue stars in  Hodge\,301 listed by \citet{Evans_2015} are Be stars, and they also account for seven of the 13 blue stars in SL\,639. If we exclude these Be stars from consideration we find that, of the remainder, ages have been published for three blue supergiants (BSGs) in each cluster \citep{McEvoy_2015} while ages for some of the non-BSG turn-off (TOF) stars have been published  by \citet{Schneider_2018}; four stars in Hodge\,301 and two stars in SL\,639. 
The mean ages of these stellar groups, as well as their age ranges are listed in Table~\ref{lb:ages}.

\begin{table}
\begin{center}
{\small
\caption{Comparison of the mean ages and age spread of identified red supergiants (RSG), blue supergiants (BSG), and non-Be, turn-off (TOF) stars in Hodge\,301 and SL\,639.}\label{lb:ages}
\begin{tabular}{lllll}
\hline\hline
Cluster & Obj. type & Number & Mean (Myr) & Spread (Myr) \\
\hline
\\
Hodge\,301 & RSG & 4 &  15 & 14--24 \\
& BSG & 3 &  12  & 9--15 \\
& TOF & 4 &  8 & 3--20 \\
\\
SL\,639 & RSG & 4 & 18 & 12--22 \\
 & BSG & 3 & 10 & 7--12 \\
 & TOF & 2 & 19 & 18--20 \\
\hline
\end{tabular}
}
\end{center}
\end{table}

If the apparent age spreads of RSGs in a given star cluster are caused by this red straggler effect, the true cluster age  would correspond simply to the age of the least luminous RSG as derived from single-star models. As shown in Fig.~\ref{HR_all}, the masses derived for our RSGs vary by less than a factor of two, and the red straggler interpretation may thus apply. In this context, the estimated age of Hodge~301 is 24$^{+5}_{-3}$\,Myr (see Table~\ref{tb:derived}), with an uncertainty according to the luminosity error of its least luminous RSG from Fig.~\ref{vfts_lage_all}. Similarly, we
estimated an age of 22$^{+6}_{-5}$\,Myr for SL\,639, although we note that this is
defined by VFTS\,852 and 2090, which are at larger radii from the cluster centre (see discussion
in Section~\ref{sec:introduction}). Referring to Table~\ref{lb:ages}, we see that the oldest stars in the RSG and TOF groups are similar, while the BSG stars are systematically younger than these limits. This picture is consistent with the oldest TOF and RSG stars more correctly reflecting the ages of these clusters, of 20--25 Myr, while the younger stars are red and blue stragglers as argued above.

The general conclusion is that each evolutionary model would predict a small luminosity range of RSGs during He-burning phase. In this way, the red stragglers as binary products are required for the explanation of the luminosity spread of RSGs in a coeval cluster.

\subsection{Further examples: RSGC1 and NGC\,2100}

\citet{Davies_2008} used the Geneva models to analyse 15 RSGs in the Galactic cluster RSGC1. They found that a single 12\,Myr age isochrone described the complete population of RSGs, with the spread in their luminosities attributed to strong mass loss and uncertainties in the estimated cluster distance. The least luminous RSG in RSGC1 has log(L/L$_{\sun}$) = 4.87$_{-0.14}^{+0.13}$. Using our technique this corresponds to an age of 13$_{-2}^{+3}$ Myr, in good agreement with the estimate from \citeauthor{Davies_2008}

At this point, an important question arises -- how should we  interpret the spread of RSG luminosities in a coeval cluster? Is it due to single-star evolution \citep[as discussed by][]{Davies_2008} or binary evolution (as suggested in the previous section)? 
As shown in Fig.~\ref{vfts_lage_all}, a single evolutionary track can indeed show a significant luminosity range during the He-burning phase, but as noted in the previous section, the early part of this
range is very rapid and observing a RSG during the early onset of the He-burning phase is small. Thus, a given sample of coeval RSGs would be expected to occupy quite a narrow range of luminosity ($\sim$0.2 dex in log(L/L$_{\sun}$)). Although single-star evolution might account for
some of the spread, we argue that the binary channel discussed above is also a potentially significant factor.


Prompted by these results and the low velocity dispersion reported by \citet{p2100}, we turned to the RSG population of the NGC\,2100 cluster in the LMC. We used results from \citet{Beasor_2016} for 18 RSGs in NGC\,2100  to construct the L--Age diagram shown in Fig.~\ref{vfts_lage_ngc2100}, giving an estimated age of 
$23^{+4}_{-2}$\,Myr for the least luminous RSG. In the figure we highlight the
five stars that are most distant from the visual cluster centre (\#1, 3, 6, 12, and 17 from
\citeauthor{Beasor_2016}), finding an estimated age of 21.5\,$\pm$\,3\,Myr for the least luminous
star in this sub-sample, and providing support that they are coeval with the main body of the cluster.
These estimates are in good agreement with the value of 20.6\,$\pm$\,1.6\,Myr derived from analysis of the
star-formation history of the cluster by \citet{Niederhofer_2015}. The reference measurements were based on all stellar populations in NGC 2100, including the main sequence stars and the evolved red stars. Thus, the agreement in the age estimates is reasonable.

The large number of RSGs in NGC\,2100 enables us to (very) roughly  estimate the red straggler fraction in an example coeval cluster. By assuming the stellar lifetime scales as $t \sim M^{-2.5}$, the relative mass range scales as ${\rm d}M/M\sim 0.4{\rm d}t/t$. If we assume that the RSG phase is only 10$\%$ of the total lifetime, the relative mass range of single RSGs in a coeval cluster should be of order 5$\%$. From the comparison in Fig.~\ref{vfts_lage_ngc2100} the least luminous RSG in NGC\,2100 has a mass of $\approx$10\,$M_{\odot}$. If we then consider that all RSGs with masses up to 10.5$M_{\odot}$ are effectively single, then eight stars in Fig.~\ref{vfts_lage_ngc2100}, 
(with L/L$_{\odot}$\,$=$\,4.43--4.56\,dex), could be single stars. The remaining 10 stars are potentially red stragglers, giving a fraction of red stragglers in this coeval cluster of $\sim$55\%. 

There are large uncertainties on this fraction, but we
conclude that this channel is potentially a significant factor in the observed populations of young, massive clusters. If the most luminous RSGs in a cluster are also those with the largest mass-loss rates \citep[e.g.][]{Beasor_2016}, the latter might also help reveal potential red stragglers.


\section{Conclusions} \label{sec:conclusion}

We have undertaken a photometric study of the RSG population of the 30~Doradus region in 
the LMC, which comprises 20 candidate RSGs. With the benefit of detailed analysis of the early-type stars from the VFTS, we used O-type stars around our cool-star sample to define the line-of-sight extinction toward each target. We estimated physical parameters for the sample, adopting the single-band technique to estimate luminosities as the most reliable approach, and we employed $J$-band photometry to mitigate the
impact of extinction (at shorter wavelengths) and possible excesses from mass loss
(at longer wavelengths). We showed that accurate correction of interstellar reddening is crucial and can not be neglected in determination of the physical parameters of RSGs in young stellar clusters. It is possible that we have underestimated the luminosities of the most luminous RSGs because of circumstellar dust  -- while this will affect the inferred age spread of a given population of RSGs, it will not influence the age estimated from the least luminous RSGs.

From analysis of the results for RSGs in the LMC from
\citet{Tabernero_2018} we present a new empirical calibration of 
T$_{\rm eff}$ vs.\ $(J-K_{\rm s})_0$ (Eqn.~\ref{Teff_relation}) to estimate temperatures
of our sample. This relation should also serve as a useful tool in future
extragalactic work where we often only have photometric information on populations
of RSGs (Patrick \& Britavskiy, in prep.). From our analysis we conclude that the sample contains 15 RSGs, 
four AGB (or low-luminosity RSG) stars, and one foreground object.

We introduced the luminosity--age diagram based on evolutionary tracks for the LMC to estimate
ages of our RSGs, finding ranges of 14 to 24\,Myr and 12 to 22\,Myr, for Hodge~301 and SL\,639, respectively. Assuming that binary mass transfer and mergers can produce more massive and luminous RSGs than expected from single-star evolution at a given age, analogous to the blue straggler phenomenon at the main sequence turn-off, we argue that the most luminous RSGs in these two star clusters are red stragglers. In this scenario, the least luminous RSGs in the clusters 
would effectively be the products of single-star evolution, and thus their ages
derived from comparisons with single-star tracks are expected to represent the cluster ages. Based on these arguments, we estimate ages of 24$^{+5}_{-3}$\,Myr for Hodge\,301, and 22$^{+6}_{-5}$\,Myr for SL\,639.
We have also applied our methods to the RSG population of NGC\,2100, finding a similarly large apparent spread in the L--Age diagram, with an estimated age of $23^{+4}_{-2}$\,Myr.

\begin{acknowledgements}
We thank the anonymous referee for helpful comments that have improved the manuscript. NB, AH, and LRP acknowledge support from grant AYA2015- 68012-C2- 1-P from the Spanish Ministry of Economy and Competitiveness (MINECO) and also acknowledge support  from the grant of Gobierno de Canarias (ProID2017010115). CJE is grateful to Rob Ivison for his sparkling motivation throughout this project. This publication makes use of data products from the Two Micron All Sky Survey, which is a joint project of the University of Massachusetts and the Infrared Processing and Analysis Center/California Institute of Technology, funded by the National Aeronautics and Space Administration and the National Science Foundation.
\end{acknowledgements}

\bibliographystyle{aa}
\bibliography{ref}

\newgeometry{margin=1.5cm}
\begin{landscape}
\begin{table*}
{\tiny
\caption{Estimated fundamental physical parameters for candidate RSGs in 30~Dor.}              
\label{tb:derived}      
\centering                                      
\begin{tabular}{clccccccccccrrl}          
\hline\hline                        
\# & VFTS ID	& T$_{\rm eff}^{(V-K)}$	&    L$^{(V-K)}$	& T$_{\rm eff}^{(J-K)}$	&	L$^{(J-K)}$ &	L$^{(I)}$ &	L$^{(J)}$ &	L$^{(K)}$ &		R	& log~$g$ & M & Age (B) &  Age (P)    & Notes \\
	          &                     	& (K)            	& log(L/L$_{\sun}$)  &  (K)          	&  log(L/L$_{\sun}$) &   log(L/L$_{\sun}$) &  log(L/L$_{\sun}$)  &  log(L/L$_{\sun}$) & (R$_{\sun}$) & log(cm$~s^{-2}$) & (M$_{\sun}$)&   (Myr)   &  (Myr)   &      \\
\hline
1	&023 	&4259   $\pm$ 222	&4.28	$\pm$ 0.09	&3873	$\pm$ 144	&4.09	$\pm$ 0.14	&4.33	$\pm$ 0.17	&4.17	$\pm$ 0.18	&4.22 $\pm$ 0.17              &268 $\pm$ 56$\phantom{0}$     &  0.45(?)   &  $\phantom{0}$7.7 $\pm$ 3.2   &33.6$_{-6}^{+10}$ &33.0$_{-8}^{+15}$  & AGB candidate  \\
2	&081 	&4269	$\pm$ 131&4.82	$\pm$ 0.05	&3935	$\pm$ 142	&4.67	$\pm$ 0.13	&4.86	$\pm$ 0.08	&4.74	$\pm$ 0.09	&4.77	$\pm$ 0.10	&507 $\pm$ 59$\phantom{0}$ & 0.3$\phantom{0}$ & 18.7 $\pm$ 4.2 	&16.1$_{-3}^{+5}$   &15.6$_{-2}^{+3}$   &  \\
3	&198	&4029	$\pm$ 210	&4.74	$\pm$ 0.09	&3884	$\pm$ 150	&4.63	$\pm$ 0.16	&4.72	$\pm$ 0.20	&4.67	$\pm$ 0.24	&4.77	$\pm$ 0.20	&495 $\pm$ 132& 0.2$\phantom{0}$& 14.7 $\pm$ 7.0&17.7$_{-5}^{+7}$&17.1$_{-5}^{+8}$&    \\
4	&222	&4084	$\pm$ 146	&4.25	$\pm$ 0.06	&3935	$\pm$ 145	&4.15	$\pm$ 0.14	&4.28	$\pm$ 0.11	&4.19	$\pm$ 0.13	&4.25	$\pm$ 0.13	&276 $\pm$ 46$\phantom{0}$ & 0.5$\phantom{0}$(?) & $\phantom{0}$8.7 $\pm$ 2.7 &32.5$_{-5}^{+10}$&31.9$_{-6}^{+12}$&  AGB candidate   \\
5	&236	&4035	$\pm$ 114	&4.58	$\pm$ 0.05	&3932	$\pm$ 142	&4.50	$\pm$ 0.13	&4.59	$\pm$ 0.07	&4.54	$\pm$ 0.09	&4.58	$\pm$ 0.08	&403 $\pm$ 55$\phantom{0}$ & 0.4$\phantom{0}$ & 15.5 $\pm$ 4.2 &21.1$_{-4}^{+5}$ &20.5$_{-4}^{+5}$& {H301 cand.}; T$_{\rm eff}$\,$=$\,3950\,$\pm$\,105\,K (T18)  \\
6	&275	&3447	$\pm$ 37$\phantom{0}$	&5.03	$\pm$ 0.02	&3677	$\pm$ 148	&4.98	$\pm$ 0.12	&4.80	$\pm$ 0.08	&5.04	$\pm$ 0.08	&5.22	$\pm$ 0.08	&824 $\pm$ 108 &  $-$0.1$\phantom{0}\phantom{-}$ & 20.3 $\pm$ 5.2 &10.7$_{-2}^{+2}$&10.8$_{-2}^{+2}$& T$_{\rm eff}$\,$=$\,3625\,K, log(L/L$_{\sun}$)=5.13 (L07) \\
7	&281	&3632	$\pm$ 59$\phantom{0}$	&4.80	$\pm$ 0.03	&3837	$\pm$ 162	&4.78	$\pm$ 0.12	&4.79	$\pm$ 0.09	&4.84	$\pm$ 0.09	&4.92	$\pm$ 0.11	&602 $\pm$ 77$\phantom{0}$ & 0.15& 18.5 $\pm$ 4.7&14.1$_{-2}^{+3}$&14.0$_{-3}^{+3}$&{Hodge~301 member}     \\
8	&289	&4110	$\pm$ 115	&4.48	$\pm$ 0.05	&3924	$\pm$ 142	&4.36	$\pm$ 0.13	&4.51	$\pm$ 0.10	&4.41	$\pm$ 0.08	&4.46	$\pm$ 0.10	&349 $\pm$ 43$\phantom{0}$& 0.4$\phantom{0}$& 11.3 $\pm$ 2.7 &24.4$_{-4}^{+4}$&24.1$_{-5}^{+5}$&{Hodge~301 member}   \\
9	&341	&3851	$\pm$ 283	&4.99	$\pm$ 0.12	&3763	$\pm$ 160	&4.88	$\pm$ 0.18	&4.87	$\pm$ 0.28	&4.94	$\pm$ 0.30	&5.04	$\pm$ 0.35	&762 $\pm$ 275 & 0.0$\phantom{0}$ & \phantom{0}20.8 $\pm$ 10.0	&12.2$_{-3}^{+7}$&12.4$_{-4}^{+7}$& T$_{\rm eff}$\,$=$\,3737\,$\pm$\,94\,K (T18)\\ 
10	&655	&4225	$\pm$ 305&4.34	$\pm$ 0.12	&3920	$\pm$ 150	&4.19	$\pm$ 0.16	&4.35	$\pm$ 0.19	&4.22	$\pm$ 0.19	&4.25	$\pm$ 0.25	&288 $\pm$ 65$\phantom{0}$ &  0.45(?) & $\phantom{0}$6.6 $\pm$ 3.7 &31.5$_{-7}^{+8}$&30.9$_{-9}^{+15}$&  AGB candidate \\
11	&744	&4108	$\pm$ 533&4.26	$\pm$ 0.21	&3827	$\pm$ 179	&4.09	$\pm$ 0.20	&4.30	$\pm$ 0.38	&4.20	$\pm$ 0.37	&4.24	$\pm$ 0.38	&298 $\pm$ 149 & 0.35(?) & $\phantom{0}$8.7$ \pm$ 6.0	&32.4$_{-13}^{+22}$&31.8$_{-13}^{+22}$& AGB candidate \\
12	&793	&5566	$\pm$ 299&2.15	$\pm$ 0.10	&4254	$\pm$ 143	&2.26	$\pm$ 0.13	&2.30	$\pm$ 0.11	&2.37	$\pm$ 0.12	&2.19	$\pm$ 0.12	&$\phantom{0}$29 $\pm$ 4$\phantom{0}\phantom{0}$  & -- & --	 &--&--& Foreground star \\
13	&828	&3860	$\pm$ 120&4.95	$\pm$ 0.05	&3883	$\pm$ 145	&4.88	$\pm$ 0.13	&4.99	$\pm$ 0.11	&4.94	$\pm$ 0.12	&5.03	$\pm$ 0.12	&664 $\pm$ 106 & 0.1$\phantom{0}$ & 21.0 $\pm$ 6.8	&12.2$_{-2}^{+3}$&12.4$_{-2}^{+3}$& {SL\,639 member}\\
14	&839	&3966	$\pm$ 124&4.81	$\pm$ 0.05	&3877	$\pm$ 145	&4.72	$\pm$ 0.12	&4.85	$\pm$ 0.11	&4.79	$\pm$ 0.12	&4.85	$\pm$ 0.14	&539 $\pm$ 92$\phantom{0}$ & 0.25 & 19.6 $\pm$ 5.7	&15.0$_{-4}^{+5}$&14.8$_{-3}^{+3}$&{SL\,639 member}\\
15	&852	&4363	$\pm$ 119	&4.64	$\pm$ 0.05	&3899	$\pm$ 140	&4.53	$\pm$ 0.02	&4.69	$\pm$ 0.08	&4.58	$\pm$ 0.08	&4.64	$\pm$ 0.08	&431 $\pm$ 54$\phantom{0}$ & 0.3$\phantom{0}$ & 13.6 $\pm$ 2.8	&19.9$_{-5}^{+5}$& 19.2$_{-5}^{+5}$& SL 639 candidate   \\
16	&2002   &3883	$\pm$ 220&5.04	$\pm$ 0.09	&3799	$\pm$ 150	&4.93	$\pm$ 0.16	&5.00	$\pm$ 0.17	&5.00	$\pm$ 0.24	&5.10	$\pm$ 0.23	&757 $\pm$ 211 & 0.0$\phantom{0}$ & \phantom{0.}21 $\pm$ 10	&11.3$_{-2}^{+7}$&11.4$_{-5}^{+8}$& T$_{\rm eff}$\,$=$\,3748\,$\pm$\,79\,K (T18)   \\
17	&2028   &3854	$\pm$ 131&4.94	$\pm$ 0.06	&3914	$\pm$ 143	&4.91	$\pm$ 0.14	&4.93	$\pm$ 0.10	&4.95	$\pm$ 0.14	&5.01	$\pm$ 0.14	&654 $\pm$ 111 & 0.1$\phantom{0}$ & 19.5 $\pm$ 6.0	&12.1$_{-2}^{+3}$&12.3$_{-2}^{+3}$&  T$_{\rm eff}$\,$=$\,4572\,$\pm$\,75\,K (T18)    \\
18	&2090   &4284	$\pm$ 151&4.60	$\pm$ 0.06	&3895	$\pm$ 143	&4.42	$\pm$ 0.14	&4.63	$\pm$ 0.10	&4.51	$\pm$ 0.12	&4.54	$\pm$ 0.13	&394 $\pm$ 61$\phantom{0}$ & 0.35 & 13.2 $\pm$ 3.9 &21.8$_{-5}^{+6}$&21.2$_{-4}^{+6}$&  {SL\,639 candidate}    \\
19	&WB97 5 &4858 $\pm$ 180   	&5.06	$\pm$ 0.06	&3834	$\pm$ 142	&4.67	$\pm$ 0.13	&4.72	$\pm$ 0.09	&4.72	$\pm$ 0.07	&4.83	$\pm$ 0.09	&522 $\pm$ 53$\phantom{0}$ & 0.2$\phantom{0}$ & 16.2 $\pm$ 3.5&16.5$_{-3}^{+3}$&16.0$_{-3}^{+3}$&{Hodge~301 member}    \\
20	&Mk 9& 3700 $\pm$ 60   	&5.13	$\pm$ 0.03	&3862	$\pm$ 150	&5.11	$\pm$ 0.13	&5.17	$\pm$ 0.09	&5.15	$\pm$ 0.09	&5.25	$\pm$ 0.08	&902 $\pm$ 45$\phantom{0}$ & $-$0.1$\phantom{0}$ & 23.0 $\pm$ 3.0 &9.4$_{-2}^{+2}$&9.4$_{-2}^{+3}$   \\
\hline                                             
\end{tabular}
\tablefoot{Published parameters in the final column are from L07 \citep{Levesque_2007} and T18 \citep{Tabernero_2018}. Typical uncertainties on logarithmic gravities are $\pm$\,0.1\,dex, with larger uncertainties for the candidate AGB stars, as indicated by the question marks. We presented two age estimates: `Age (B)', from evolutionary tracks from \citet{Brott2011}; `Age(P)', using the tracks from \citet{Bressan_2012}.}
}
\end{table*}
\end{landscape}
\restoregeometry

\onecolumn
\begin{appendix}
\section{Additional cool stars}
{\footnotesize
\begin{longtable}{lccccl}
\caption{Observational parameters of stars with FLAMES spectroscopy but previously omitted from the VFTS catalogue.}\label{rejects}\\
\hline
\hline
Star & $\alpha$\,(J2000) & $\delta$\,(J2000) & $V$ & $B-V$ & Ref. \\
\hline
\endfirsthead
\caption[]{\it{continued}}\\
\hline
Star & $\alpha$\,(J2000) & $\delta$\,(J2000) & $V$ & $B-V$ & Ref. \\
\hline
\endhead
\hline
\multicolumn{6}{r}{\it{continued on next page}}\\
\endfoot
\hline
\endlastfoot
2001 & 05 36 51.11 & $-$69 05 12.26 & 16.63 & 0.67 & W \\
2002 & 05 37 13.50 & $-$69 08 34.65 & 14.21 & 2.07 & Z \\
2003 & 05 37 15.24 & $-$69 00 27.91 & 13.76 & 0.86 & W \\
2004 & 05 37 22.85 & $-$69 01 49.26 & 13.74 & 0.56 & Z \\
2005 & 05 37 23.91 & $-$69 04 40.51 & 16.72 & 0.89 & W \\
2006 & 05 37 29.16 & $-$69 02 10.57 & 16.62 & 0.69 & W \\
2007 & 05 37 31.07 & $-$69 10 23.94 & 15.33 & 0.85 & W \\
2008 & 05 37 32.72 & $-$68 58 23.02 & 16.53 & 1.65 & Z \\
2009 & 05 37 33.12 & $-$69 01 57.44 & 16.79 & 1.58 & W \\
2010 & 05 37 34.60 & $-$69 12 20.59 & 16.83 & 1.04 & W \\
2011 & 05 37 39.78 & $-$69 10 12.13 & 16.86 & 1.00 & W \\
2012 & 05 37 42.10 & $-$69 09 01.54 & 16.37 & 1.78 & W \\
2013 & 05 37 44.02 & $-$69 10 35.86 & 11.94 & 0.42 & C \\
2014 & 05 37 44.65 & $-$69 00 55.54 & 16.74 & 1.32 & Z \\
2015 & 05 37 46.47 & $-$69 11 17.00 & 14.58 & 0.68 & W \\
2016 & 05 37 48.04 & $-$69 10 02.44 & 16.06 & 0.81 & W \\
2017 & 05 37 49.38 & $-$69 06 13.29 & 16.13 & 1.50 & W \\
2018 & 05 37 49.47 & $-$69 00 02.94 & 15.90 & 0.91 & W \\
2019 & 05 37 50.13 & $-$69 13 34.36 & 16.74 & 1.50 & W \\
2020 & 05 37 50.18 & $-$69 04 24.47 & 11.87 & 0.95 & C \\
2021 & 05 37 50.68 & $-$69 12 48.63 & 13.52 & 0.60 & Z \\
2022 & 05 37 51.79 & $-$69 08 08.70 & 15.63 & 0.93 & W \\
2023 & 05 37 52.38 & $-$69 00 16.09 & 15.18 & 0.76 & W \\
2024 & 05 37 53.17 & $-$69 11 51.01 & 13.88 & 0.92 & Z \\
2025 & 05 37 54.00 & $-$68 57 19.76 & 16.32 & 0.96 & Z \\
2026 & 05 37 54.63 & $-$68 58 19.48 & 15.06 & 0.96 & W \\
2027 & 05 37 58.26 & $-$69 02 09.45 & 16.57 & 0.87 & W \\
2028 & 05 37 58.67 & $-$69 14 24.07 & 13.27 & 2.15 & Z \\
2029 & 05 38 02.51 & $-$69 03 41.98 & 16.04 & 0.75 & W \\
2030 & 05 38 03.93 & $-$69 09 27.03 & 15.83 & 0.66 & W \\
2031 & 05 38 04.96 & $-$69 07 34.28 & 15.30 & 0.96 & W \\
2032 & 05 38 10.34 & $-$69 15 08.35 & 16.32 & 0.68 & W \\
2033 & 05 38 13.18 & $-$69 05 36.59 & 14.25 & 0.78 & Z \\
2034 & 05 38 14.38 & $-$69 06 04.97 & 16.36 & 1.01 & W \\
2035 & 05 38 14.64 & $-$69 00 57.71 & 16.34 & 2.31 & W \\
2036 & 05 38 14.72 & $-$69 14 52.09 & 16.81 & 1.69 & W \\
2037 & 05 38 16.03 & $-$68 58 03.82 & 14.87 & 0.72 & W \\
2038 & 05 38 17.85 & $-$69 15 37.68 & 16.74 & 1.50 & W \\
2039 & 05 38 19.11 & $-$68 59 01.96 & 16.26 & 0.68 & W \\
2040 & 05 38 19.87 & $-$68 56 27.17 & 14.50 & 0.68 & W \\
2041 & 05 38 23.37 & $-$68 59 57.72 & 16.78 & 0.61 & W \\
2042 & 05 38 23.64 & $-$69 14 57.09 & 16.86 & 0.65 & W \\
2043 & 05 38 24.92 & $-$69 11 13.82 & 16.04 & 0.53 & W \\
2044 & 05 38 26.50 & $-$69 03 11.08 & 16.33 & 0.75 & P (P93-89) \\
2045 & 05 38 30.05 & $-$69 06 25.94 & 16.26 & 1.08 & S (S99-500) \\
2046 & 05 38 31.02 & $-$69 01 15.89 & 12.00 & 0.49 & P (P93-9009) \\
2047 & 05 38 35.29 & $-$69 03 54.15 & 15.61 & 0.72 & P (P93-459) \\
2048 & 05 38 38.52 & $-$69 06 46.57 & 15.37 & 0.62 & S (S99-222) \\
2049 & 05 38 39.22 & $-$69 15 30.38 & 16.32 & 0.65 & W \\
2050 & 05 38 41.19 & $-$69 08 51.93 & 16.89 & 1.44 & W \\
2051 & 05 38 43.14 & $-$69 08 35.59 & 15.35 & 0.61 & W \\
2052 & 05 38 43.92 & $-$68 59 40.06 & 16.68 & 0.44 & W \\
2053 & 05 38 43.92 & $-$69 12 43.49 & 16.73 & 1.47 & W \\
2054 & 05 38 45.49 & $-$69 00 17.30 & 13.89 & 0.69 & W \\
2055 & 05 38 48.13 & $-$69 09 45.13 & 16.73 & 0.70 & W \\
2056 & 05 38 48.22 & $-$69 08 05.65 & 15.02 & 0.60 & P (P93-1428) \\
2057 & 05 38 52.56 & $-$69 11 24.48 & 16.67 & 2.06 & W \\
2058 & 05 38 54.69 & $-$69 07 44.79 & 12.32 & 0.97 & P (P93-1684) \\
2059 & 05 38 57.28 & $-$69 00 33.62 & 15.89 & 0.69 & W \\
2060 & 05 38 58.53 & $-$68 58 03.19 & 14.77 & 0.60 & W \\
2061 & 05 39 00.48 & $-$69 08 41.27 & 15.00 & 0.67 & W \\
2062 & 05 39 02.32 & $-$69 14 59.62 & 15.63 & 0.62 & W \\
2063 & 05 39 03.53 & $-$69 14 14.43 & 16.86 & 0.60 & W \\
2064 & 05 39 07.41 & $-$69 04 20.35 & 14.90 & 0.60 & W \\
2065 & 05 39 12.46 & $-$68 59 03.83 & 16.40 & 0.53 & W \\
2066 & 05 39 12.51 & $-$69 04 08.84 & 14.61 & 0.76 & W \\
2067 & 05 39 13.62 & $-$69 14 38.03 & 16.91 & 0.67 & W \\
2068 & 05 39 14.93 & $-$69 11 51.01 & 14.14 & 0.77 & Z \\
2069 & 05 39 18.18 & $-$69 08 48.31 & 15.02 & 0.67 & W \\
2070 & 05 39 24.64 & $-$69 02 19.31 & 16.79 & 0.70 & W \\
2071 & 05 39 26.26 & $-$69 00 19.39 & 16.87 & 0.65 & W \\
2072 & 05 39 26.58 & $-$69 06 04.75 & 16.12 & 0.83 & W \\
2073 & 05 39 27.65 & $-$69 00 26.64 & 15.93 & 0.76 & W \\
2074 & 05 39 30.26 & $-$69 06 36.36 & 15.93 & 0.73 & W \\
2075 & 05 39 32.80 & $-$69 00 07.39 & 16.49 & 1.11 & Z \\
2076 & 05 39 33.63 & $-$69 12 23.50 & 13.64 & 0.64 & Z \\
2077 & 05 39 35.49 & $-$69 04 38.72 & 16.93 & 1.46 & W \\
2078 & 05 39 38.09 & $-$69 09 14.59 & 15.28 & 0.97 & W \\
2079 & 05 39 39.11 & $-$68 59 13.17 & 16.88 & 0.63 & W \\
2080 & 05 39 39.21 & $-$69 07 01.98 & 16.55 & 0.83 & W \\
2081 & 05 39 41.79 & $-$69 11 48.87 & 16.63 & 0.91 & W \\
2082 & 05 39 44.70 & $-$69 04 30.18 & 15.57 & 1.31 & W \\
2083 & 05 39 47.83 & $-$69 11 39.14 & 16.25 & 0.52 & W \\
2084 & 05 39 53.07 & $-$69 08 50.06 & 16.61 & 1.07 & W \\
2085 & 05 39 57.71 & $-$69 06 36.28 & 14.17 & 0.77 & Z \\
2086 & 05 39 58.11 & $-$69 02 40.95 & 16.52 & 1.78 & W \\
2087 & 05 40 00.16 & $-$69 02 23.29 & 14.09 & 0.99 & W \\
2088 & 05 40 00.77 & $-$69 01 39.95 & 16.82 & 1.67 & W \\
2089 & 05 40 03.82 & $-$69 03 24.46 & 16.61 & 1.64 & W \\
2090 & 05 40 07.01 & $-$69 11 41.50 & 14.96 & 1.98 & W \\
2091 & 05 40 08.27 & $-$69 08 29.90 & 16.82 & 1.66 & W \\
2092 & 05 40 09.45 & $-$69 02 54.60 & 16.72 & 1.34 & W \\
2093 & 05 40 11.44 & $-$69 11 48.81 & 16.94 & 1.51 & W \\
2094 & 05 40 13.62 & $-$69 05 51.10 & 16.71 & 0.74 & W \\
2095 & 05 40 13.63 & $-$69 06 21.75 & 16.65 & 1.08 & W \\
2096 & 05 40 14.28 & $-$69 02 09.56 & 16.76 & 1.14 & W \\
2097 & 05 40 14.42 & $-$69 06 41.55 & 16.88 & 1.66 & W \\
2098 & 05 40 24.75 & $-$69 02 24.25 & 16.20 & 0.63 & W \\
2099 & 05 40 26.88 & $-$69 08 20.84 & 16.82 & 1.61 & W \\
2100 & 05 40 28.05 & $-$69 07 48.54 & 16.58 & 0.65 & W \\
2101 & 05 40 28.99 & $-$69 03 46.24 & 14.50 & 0.70 & W \\
2102 & 05 40 33.41 & $-$69 05 57.69 & 16.68 & 1.65 & W \\
\hline
\end{longtable}
\tablefoot{Sources of photometry are (in order of preference where available): S \citep{s99}, W (WFI, Paper~I), P \citep{p93}, Z \citep{mcps}, C (CTIO, Paper~I).}
}

\end{appendix}

\end{document}